\documentclass[aps,twocolumn,showpacs,superscriptaddress,groupedaddress,nofootinbib]{revtex4}  
\usepackage{graphicx}  
\usepackage{dcolumn}   
\usepackage{bm}        
\usepackage{amssymb}   
\usepackage{amsmath}
\usepackage{mathtools}
\usepackage{comment}
\usepackage{aas_macros}
\usepackage{endnotes}
\usepackage{color}
\usepackage[bookmarks=true, colorlinks=true]{hyperref}
\usepackage{soul}
\usepackage[normalem]{ulem}

\newcommand{\be}{\begin{equation}}
\newcommand{\ee}{\end{equation}}
\newcommand{\beq}{\begin{eqnarray}}
\newcommand{\eeq}{\end{eqnarray}}

\newcommand{\n}{neutrino}

\newcommand{\hk}{Hyper-K}

\definecolor{mypink1}{rgb}{0.858, 0.188, 0.578}
\definecolor{mygreen}{rgb}{0, 0.4, 0.07}
\definecolor{mypurple}{rgb}{0.5, 0, 0.85}

\begin{document}

\title{Detectability of neutrino-signal
fluctuations induced by the hadron-quark phase transition in failing
core-collapse supernovae}

\author{Zidu Lin }
\affiliation{Department of Physics and Astronomy, University of Tennessee Knoxville, Knoxville 37996, US}

\author{Shuai Zha}
\affiliation{Tsung-Dao Lee Institute, Shanghai Jiao Tong University, Shanghai 200240, China}
\affiliation{Yunnan Observatories, Chinese Academy of
Sciences (CAS), Kunming 650216, China}
\affiliation{Key Laboratory for the Structure and Evolution of Celestial Objects, CAS, Kunming 650216, China}
\affiliation{International Centre of Supernovae, Yunnan Key Laboratory,
Kunming 650216, China}

\author{Evan P. O’Connor}
\affiliation{The Oskar Klein Centre, Department of Astronomy, Stockholm University, AlbaNova, SE-106 91 Stockholm, Sweden}

\author{Andrew W. Steiner}
\affiliation{Department of Physics and Astronomy, University of Tennessee Knoxville, Knoxville 37996, US}
\affiliation{Physics Division, Oak Ridge National Laboratory, Oak Ridge 37831, US}

\date{\today}

\begin{abstract}
We introduce a systematic and quantitative methodology for establishing the presence of neutrino oscillatory signals due to the hadron-quark phase transition (PT) in failing core-collapse supernovae from the observed neutrino event rate in water- or ice-based neutrino detectors. The methodology uses a likelihood ratio in the frequency domain as a test-statistic; it is employed for quantitative analysis of neutrino signals without assuming the frequency, amplitude, starting time, and duration of the PT-induced oscillations present in the neutrino events and thus it is suitable for analyzing neutrino signals from a wide variety of numerical simulations. We test the validity of this method by using a core-collapse simulation of a 17 solar-mass star by Zha \emph{et al.} (2021) \cite{Zha:2021fbi}. Based on this model, we further report the presence of a PT-induced oscillations quantitatively for a core-collapse supernovae out to a distance of $\sim 10$ kpc, $\sim 5$ kpc for IceCube and to a distance of $\sim 10$ kpc, $\sim 5$ kpc and $\sim 1$ kpc for a 0.4 Mt mass water Cherenkov detector. This methodology will aid the investigation of a future galactic supernova and the study of hadron-quark phase in the core of core-collapse supernovae.

%
\end{abstract}

\maketitle

\section{Introduction}
In 1987 the neutrino burst from supernova (SN) 1987A in the Large Magellanic Cloud was observed within a time interval of about 15 seconds, with a total of 24 neutrino events detected in KAMIOKANDE-II, IMB and Baksan detectors \cite{1987PhRvL..58.1490H,Bionta:1987qt,ALEXEYEV1988209}. Despite the small number of observed neutrinos, the mean energy and temporal spreading of these core-collapse SN (CCSN) neutrinos helped to pave the way for a new era of neutrino and CCSNe physics \cite{1987PhRvL..58.1490H,Page:2020gsx,1989ApJ...340..426L}. In the last three decades, the capabilities for CCSN neutrino detection have increased by orders of magnitude. The existing and the next generation neutrino detectors will enable the observation of neutrinos from the next galactic CCSN with good statistics, and thus provide unprecedented and detailed features of the neutrino signal \cite{Scholberg:2012id}. What can we learn from the next CCSN neutrino detection? Since neutrinos are only interacting with matter via weak interactions, they can be sensitive probes of nuclear physics \cite{Muller:2019upo,Zha:2020gjw, Warren:2019lgb, Sagert:2008ka,Horiuchi:2018ofe} and hydrodynamics \cite{Tamborra:2013laa,Tamborra:2014aua,Nagakura:2020qhb,Horiuchi:2018ofe} in the inner region of CCSNe. 

The phase transition (PT) from hadronic to deconfined quark matter is being explored by both terrestrial experiments such as heavy-ion experiments \cite{Busza:2018rrf} and observational studies of compact objects \cite{Most:2021zvc}. In the core of CCSNe and in the interior of neutron stars, the densities easily exceed the saturation density of 
$n_0\approx 0.16~\mathrm{fm}^{-3}$, which naturally raises a question: Does the core of compact objects contain non-leptonic degrees of freedom other than neutrons and protons? In search for the quark matter in compact stars, there are mainly three scenarios being explored: (1) quark matter in old accreting neutron stars \cite{Lin_2006, Abdikamalov:2008df, Niebergal:2010ds}, (2) in protoneutron stars (PNS) after the CCSNe explosions \cite{Pons:2001ar, Keranen:2004vj} and (3) in protocompact stars (PCS) produced by progenitor's iron cores in early postbounce phase of CCSNe \cite{Sagert:2008ka}. In the second scenario, the metastability and subsequent collapse to a black hole of a PNS containing quark matter are observable in current and planned neutrino detectors \cite{Pons:2001ar}. Additionally, neutrinos in a quark nova may deposit sufficient energy in the outer layers of the star to cause significant mass ejection of the nuclear envelope, which could be relevant to r-process products injected into the inter-stellar medium \cite{Keranen:2004vj}.In the third scenario, the PT may soften the equation of state (EoS), leading to further collapse and second bounce of these PCS cores \cite{Sagert:2008ka,Zha:2020gjw,Zha:2021fbi, Fischer:2017lag}, in many cases dramatically altering the ensuing dynamics. Additionally, since in this scenario the PT appears in the early postbounce phase, imprints of PT on both the neutrino and gravitational-wave (GW) signals are expected \cite{Sagert:2008ka,Zha:2020gjw,Zha:2021fbi,Kuroda:2021eiv}.   

In the recent work by \emph{Zha et al.} \cite{Zha:2021fbi}, it was found, for their choice of EoS, that for a bounce compactness parameter of $0.24<\xi_{2.2}<0.51$ (see the definition of compactness $\xi$ in section \ref{ssec:PTphysics}), the PCS experiences a second bounce with subsequent oscillations. This induces oscillatory neutrino signals that could be observed on Earth. The oscillatory signals observed in the \emph{Zha et al.} models start at $\sim 500-1000~ \mathrm{ms}$ after bounce and last for $\sim 50~ \mathrm{ms}$, with a frequency at $\sim 800 ~\mathrm{Hz}$. As shown in the numerical simulations, the damping time and the oscillation frequency of the PT-induced neutrino signals are closely related to the oscillation and damping of the PCSs in the failing CCSNe cores. The oscillation and the damping of compact objects has received increasing attention since it may (1) alter the GW signals of binary mergers in post merger phase \cite{Alford:2019qtm} and (2) probe the nuclear physics such as properties of bulk viscosity of hot and dense matter in the interior of massive stars \cite{Most:2021zvc}. In this work, we explore the neutrino counterpart that probes the hot and dense matter properties in compact objects, and show that besides GW probes, the existence and properties of a hadron-quark phase in the compact core of CCSNe can be verified and measured in a systematic and quantitative way using neutrino signals from a future galactic CCSN.

The paper is outlined as follows: In section \ref{sec:general}, we introduce the essential physics of CCSN neutrino detection and the hadron-quark PT in PCSs . In section \ref{sec:maxlik}, we discuss a novel statistical method used in this work to quantitatively study the existence and the oscillation features of PT-induced neutrino signals . In section \ref{sec:result}, we provide our statistical analysis results by using the neutrino signal from models by \emph{Zha et al.} as a test bed. Additionally, we suggest a method to constrain the effective bulk viscosity of the hadron-quark phase by measuring the frequency and the duration of the PT-induced neutrino signal. We then discuss the potential importance of the physical bulk viscosity in future numerical simulations. Finally, we conclude in section \ref{sec:disc}.  

\section{Generalities}
\label{sec:general}
 
\subsection{Supernova neutrino detection}
In this section, we briefly discuss neutrino detection methodologies in tank-based water-Cherenkov detectors and in long-string water-Cherenkov detectors such as IceCube. We further discuss the performance of observing MeV galactic CCSN neutrinos in these two types of neutrino detectors.

In tank-based water-Cherenkov detectors, for example Super-K \cite{2007ApJ...669..519I} or Hyper-K \cite{Abe:2018uyc}, the main channel for CCSN neutrino detection is inverse beta decay (IBD) $\bar{\nu}_e+p\rightarrow e^++n$. In this channel, individual positrons are detected via their Cherenkov photon signal with high efficiency and excellent time resolution (microseconds or less). In this way, a CCSN neutrino event triggered in a tank-based water-Cherenkov detector corresponds to an individual neutrino IBD reaction within the volume of the detector. Both the $\bar{\nu}_e$ energy and the flux intensity could be measured with this type of detector. In IceCube, $5160$ photomultiplier tubes were placed in the deep Antarctic ice to detect particle induced photons, with the main goal of detecting neutrinos with energies greater than $100~\mathrm{GeV}$. For a burst of MeV neutrinos, e.g. from a CCSN, Cherenkov light induced by neutrino interactions increase the count rate of all light sensors above their average value. Although this increase in an individual light sensor is not statistically significant, the effect will be clearly seen when considered collectively over the many sensors. Therefore, individual IBD reactions can not be resolved and the $\bar{\nu}_e$ energy can not be measured in IceCube, but MeV neutrinos can still be detected en masse in IceCube due to a collective rise in all photomultiplier rates on top of the dark noise. 

The performance of neutrino detectors mainly depends on two quantities: (1) the detector size and (2) the background noise in the detector. Next generation water Cherenkov detectors, like the planned Hyper-K, consist of megaton-scale water tanks and are expected to see $\sim 10^5$ neutrino events from a CCSN at 10 kpc through the IBD channel. The background events in deep underground tank-based water Cherenkov detectors are mainly due to other neutrino sources such as cosmic rays or detector impurities, and generally are negligible over the short timescale ($\sim$10s of seconds) of a galactic CCSN. In this work, a water Cherenkov detector of $0.37\,\mathrm{Mt}$\footnote{The first water-Cherenkov detector of Hyper-Kamiokande, with the total mass of 258 $\mathrm{kt}$, is expected to start operations in 2027~\cite{Hyper-Kamiokande:2020aij}.}, in accordance with the strategy of future Hyper-Kamiokande experiments including two water-Cherenkov detectors \cite{Abe:2018uyc}, is considered and the low background events are ignored. The IceCube detector, which monitors a volume of $1~\mathrm{km}^3$ in deep Antarctic ice for neutrino signals, is an even bigger Cherenkov detector than Hyper-K, and $\sim 10^6$ neutrino interactions from a CCSN at 10 kpc will occur within the effective volume \cite{2011A&A...535A.109A,Kopke:2017req}. In our work, for IceCube, a background noise component is added using a modified Gaussian distribution taking the spread to be $1.3~\sqrt{\mu}$ where $\mu$ is the average detector background rate equal to $1475$ $\mathrm{ms}^{-1}$. The factor of 1.3 is to account for correlated hits from muons \cite{2011A&A...535A.109A,Kopke:2017req,Segerlund:2021dfz}.


\subsection{PT: physics and numerical predictions} \label{ssec:PTphysics}
Here we summarize the essential physics for the oscillatory neutrino signals from failing CCSNe and refer readers to Ref.~\cite{Zha:2021fbi} for the details. According to the current understanding of the EoS for hot and dense matter \cite{2017RvMP...89a5007O}, a hadron-quark PT may take place in the PCS as it grows in mass due to accretion. The PT can lead to the collapse of the PCS and a second bounce shock wave is launched when the core stiffens due to a high quark fraction. For a very compact progenitor, this bounce shock cannot overcome the ram pressure of the envelope and falls back onto the core. The PCS starts to oscillate with the excess kinetic energy until damped due to neutrino emission and bulk viscosity. If the PCS mass exceeds the maximum value allowed by the EoS, it collapses into a black hole.

\begin{figure}
    \centering
    \includegraphics[width=0.49\textwidth]{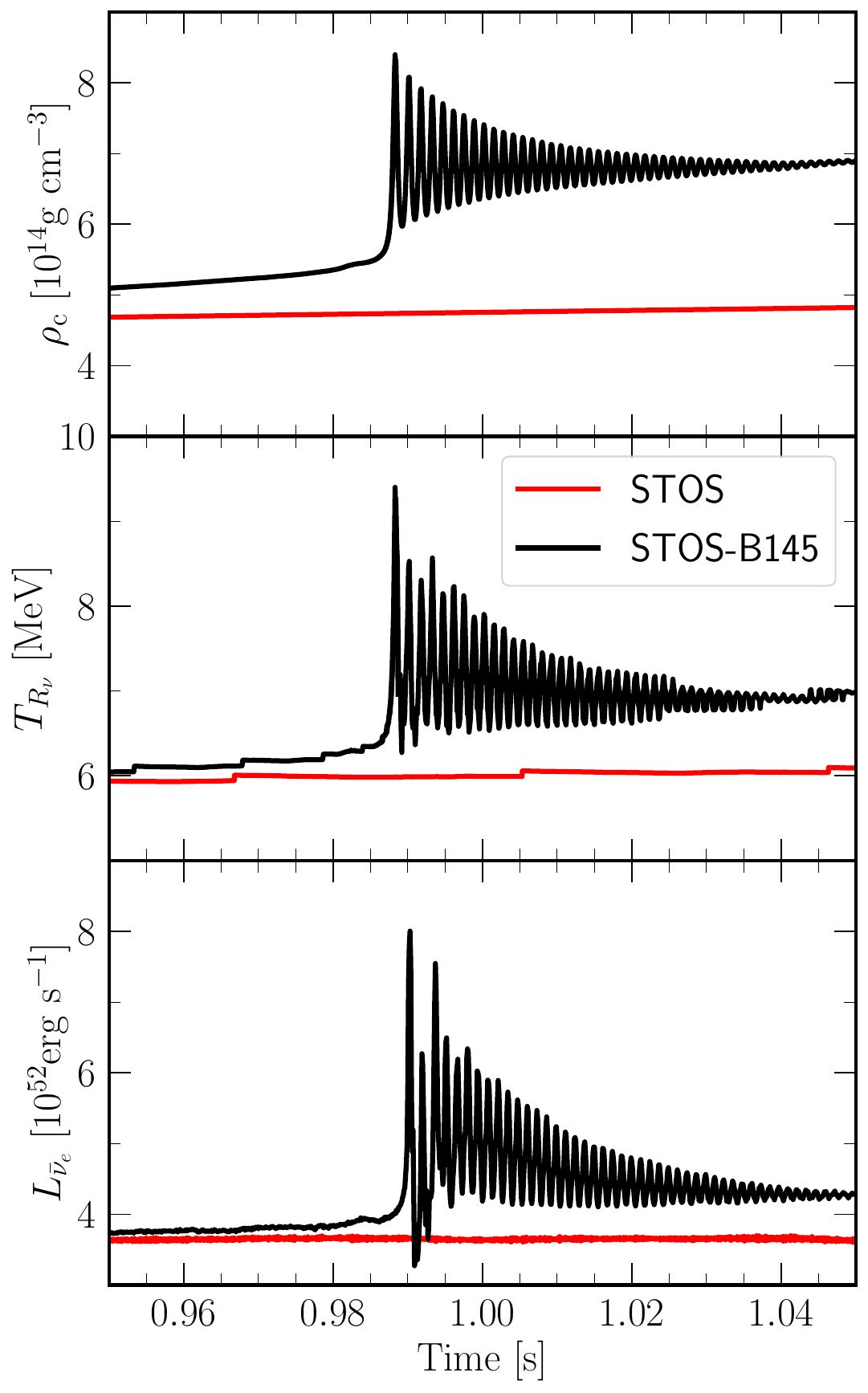}
    \caption{Central density (top), effective neutrinosphere temperature of $\bar{\nu}_e$ (middle) and luminosity of  $\bar{\nu}_e$ (bottom) as a function of time in CCSN simulations with STOS (solid red) and STOS-B145 (solid black) EoS. The zig-zag feature in $T_{R_\nu}$ is due to the finite discretization of simulation grid.}
    \label{fig:sim}
\end{figure}

\begin{figure*}[htp]
	\centering
	\includegraphics[width=0.45\textwidth]{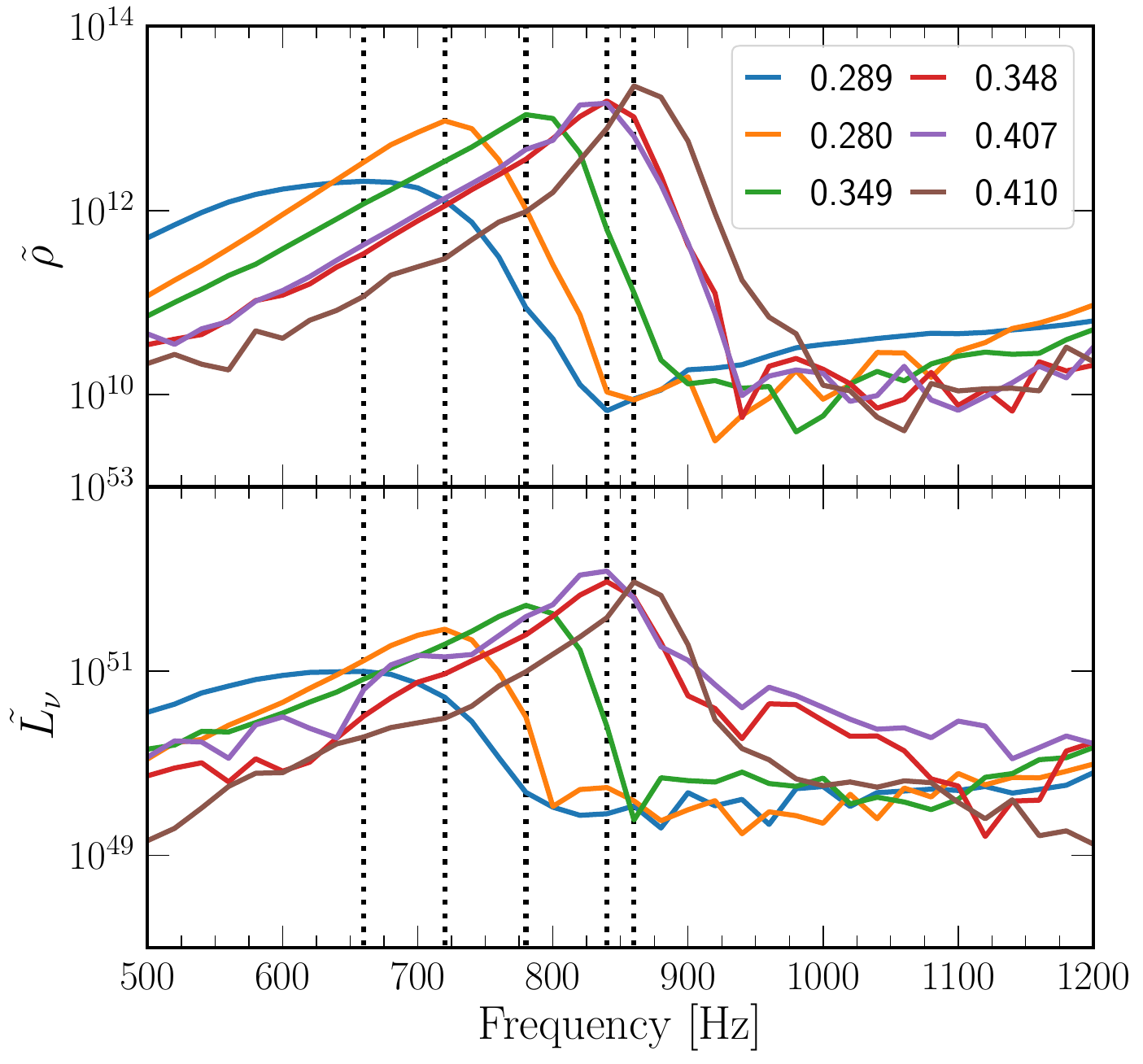}
	\caption{The upper panel shows the central density oscillation spectrum in models with different compactness parameter $\xi_{2.2}$ as marked by the legends, while the lower panel shows the oscillatory neutrino luminosity spectrum in models with different compactness. The vertical dotted lines indicate the peak frequencies in the spectrum of density oscillation.} 
	\label{fig:neurouosc}
\end{figure*}

Ref.~\cite{Zha:2021fbi} investigated the consequences of the hadron-quark PT in failing CCSNe with a hybrid EoS (STOS-B145) including hadrons and quarks \cite{2011ApJS..194...39F}. Twenty-one progenitor models were chosen from Ref.~\cite{2018ApJ...860...93S} to study the dependence of the PT on the progenitor compactness $\xi_{M}$, defined as \cite{2011ApJ...730...70O}
\begin{equation}
	\xi_M=\frac{M/M_\odot}{R(M_{\rm baryon}=M)/1000~{\rm km}}\bigg|_{t=t_0}. \label{eq:xi}
\end{equation}
Here, $R$ is the radius that encloses a baryonic mass of $M$ at a given time $t_0$. In Ref.~\cite{Zha:2021fbi}, $M$ is chosen to be $2.2~M_\odot$ which is approximately the maximum baryonic mass of PCS determined by the STOS-B145 EoS and $t_0$ is set as the time of first bounce due to the stiffening of the hadronic EoS. 

With the STOS-B145 EoS, the PCS collapses when its central density exceeds $\sim2\times\rho_{\rm sat}$. In the models explored, this occurs as early as $\sim$350\,ms after the first bounce. However, these models collapse directly to a black hole and do not have a second bounce. For a set of progenitor models with $\xi_{2.2}$ between 0.24 and 0.51, the PCS does have a second bounce and the remnant formed after this PT oscillates for $\sim$50~ms with a frequency around $\sim$1\,kHz. The oscillation of PCS leads to oscillations in the effective neutrino sphere temperature ($T_{R_{\nu}}$) with the same frequency. As shown in the bottom panel of Fig.~\ref{fig:sim}, the emitted electron anti-neutrino luminosity acquires a similar oscillatory behavior. While the neutrino transport is handled self-consistently within the simulation, we remark that the imprint of PT on the neutrino luminosity could be understood, to zeroth order, from simple oscillations in the temperature of a black body. Consequently, the luminosity scales with the temperature to the power of 4. The amplitude as well as the frequency of the oscillation of the  neutrino luminosity depends on $\xi_{2.2}$, as shown in Fig.~\ref{fig:neurouosc}. In this work, we take the model s17 (a solar-metallicity 17 $M_\odot$ progenitor from Ref.~\cite{2018ApJ...860...93S} with $\xi_{2.2}=0.349$ and with a moderate oscillatory amplitude) as a fiducial example to demonstrate its detectability with our novel statistical method. Meanwhile, we take the neutrino luminosity curve with the same progenitor model but produced using the STOS EoS \cite{2011ApJS..197...20S} that has no PT as a reference model. The evolution of central density, $T_{R_{\nu}}$ and luminosity of electron anti-neutrino as a function of time are shown together in Fig.~\ref{fig:sim} for both simulations with (STOS-B145) and without (STOS) the PT. 

\section{Testing for PT: likelihood ratio method}
\label{sec:maxlik}
In this section, we introduce a statistical method to estimate the detectability of the PT, and analyze the oscillatory features of PT-induced neutrino signals. This method is generalized from the work analyzing the features of standing accretion shock instability (SASI) by Lin \emph{et al.} (2020)\cite{Lin:2019wwm}, and features in providing quantitative results of the detectability of oscillations in neutrino events in a terrestrial Cherenkov detector, where the noise is mainly driven by the statistical fluctuations, and thus is strongly correlating with the signal itself.   

For CCSNe neutrino detection in a Cherenkov detector the recorded events are analyzed after an initial time $t_{\mathrm{ST}}$, in time bins of width $\Delta$. The  $j$-th time bin corresponds to the time $t_j=t_{\mathrm{ST}}+j\Delta$. We perform a discrete Fourier transform of the time series of neutrino events $\{N(t_j) \}$ over the time interval $[t_{\mathrm{ST}},t_{\mathrm{ST}}+ \tau ]$, where the analyzed time series starts at $t_{\mathrm{ST}}$ with duration of $\tau$ and contains $N_{\mathrm{bins}}=\tau/\Delta$ time bins. The neutrino events $\{N(t_j) \}$ were calculated using SNEWPY \cite{2021JOSS....6.3772B,2022ApJ...925..107B}.  Note that $t_{\mathrm{ST}}$ and $\tau$ are tunable parameters when searching for PT-induced signals. The discrete frequency resolution is $\delta =1/\tau$. We define the discrete Fourier-transformed \n\ signal as
\begin{equation}
h(k\delta )=\sum_{j=0}^{N_{\mathrm{bins}}-1}N(t_j)e^{i2\pi j \Delta k\delta }~,
\label{eq:h}
\end{equation}
and the one-sided power spectrum as:
\begin{equation}
 P(k\delta )=\begin{cases}
2|h(k\delta )|^2/N_{\mathrm{bins}}^2~~~ \text{for } 0<k\delta <f_{\mathrm{Nyq}}~,\\
\\
| h(k\delta )|^2/N_{\mathrm{bins}}^2 ~~~\text{for } k\delta =0. ~
\end{cases} 
\label{eq:power1}
\end{equation}
We find the probability density distribution function (PDF) of power $P(k\delta)$ follows:
\begin{equation}
\label{eq:powerProb}
	\mathrm{Prob}({P}(k\delta))=\frac{1}{2}e^{-\frac{1}{2}\left(\frac{{P}(k\delta)}{C\sigma^2}+\lambda'\right)}I_0 \left(\sqrt{\lambda'\frac{P(k\delta)}{C\sigma^2}} \right)\frac{1}{C\sigma^2},
\end{equation}
where $C=2/N_{\mathrm{bins}}^2$, $N_{\mathrm{tot}}=\sum N(t_j)$, $h_R=\sum_{j=0}^{N_{\mathrm{bins}}-1}{N(t_j)}\cos(2\pi t_j k\delta)$, $h_I=\sum_{j=0}^{N_{\mathrm{bins}}-1}{N(t_j)}\sin(2\pi t_j k\delta)$, $\lambda'=(h_R^2+h_I^2)/\sigma^2$ and $\mathrm{I}_0$ is the modified Bessel function of the first kind. Note that for Hyper-K $\sigma^2=N_{\mathrm{tot}}/2$, while for IceCube $\sigma^2=(N_{\mathrm{tot}}+1.3^2\mu)/2$. The detailed derivation of eq. \ref{eq:powerProb} is given in the Appendix of Ref. \cite{Lin:2019wwm}. To test the existence of PT imprints on neutrino signals, we first design two parametric templates which characterize the main features of neutrino signals with and without the PT-induced oscillations. For the signals with PT-induced oscillatory activity we choose a single frequency function:
\begin{equation}
N_{\mathrm{PT}}(t)=(A-n)(1+a \sin(2\pi  f_{\mathrm{PT}} t))+n~, 
\label{eq:mod2}
\end{equation}
where $A$ is the time-averaged event number per bin in the detector including instrumental noise (after possible experimental cuts), $a$ is the relative PT oscillation amplitude, $n$ is the mean value of the background events per bin ($n=0$ for \hk),  and $f_{\mathrm{PT}}$ is the nominal frequency of the PT-induced neutrino event rate oscillation frequency. 
The second template,  for the signals without PT, is a constant: 
\begin{equation}
N_{\mathrm{noPT}}(t)=A~.
\label{eq:mod0}
\end{equation}
We then Fourier transform the time series of PT and no-PT template in eq. \ref{eq:mod0} and \ref{eq:mod2} respectively, and get the power of the templates $P_{\mathrm{PT}}$ and $P_{\mathrm{noPT}}$ using eq. \ref{eq:power1}. The PDF of the power of the templates follows eq. \ref{eq:powerProb}. In the following, we denote the PDF of power of neutrino signal based on PT/noPT template as $\mathrm{Prob}(P,\Omega_{\mathrm{PT/noPT}})$, where $\Omega_{\mathrm{PT/noPT}}$ is the set of parameters in PT(no-PT) template (see eq. \ref{eq:mod2} and eq. \ref{eq:mod0}). Given the parametric PT and no-PT template, we then define the likelihood functions of neutrino signals. The likelihood that the neutrino power spectrum $\{P(k\delta)\}$ agrees with PT template is:
\begin{equation}
L(\{\mathcal{P}(k\delta)\},\Omega_{\mathrm{PT}})=\prod_{k=\lfloor400 \mathrm{Hz}/\delta \rfloor}^{\lfloor950 \mathrm{Hz}/\delta\rfloor}\mathrm{Prob}(\mathcal{ P}(k\delta), \Omega_{\mathrm{PT}})~,
\label{eq:likeliPT}
\end{equation}
while the likelihood that neutrino power spectrum $\{P(k\delta)\}$ agrees with no-PT template is:
\begin{equation}
L(\{\mathcal{ P}(k\delta)\},\Omega_{\mathrm{noPT}})=\prod_{k=\lfloor400 \mathrm{Hz}/\delta \rfloor}^{\lfloor950 \mathrm{Hz}/\delta\rfloor}\mathrm{Prob}(\mathcal{ P}(k\delta), \Omega_{\mathrm{noPT}})~.
\label{eq:likelinoPT}
\end{equation}
We then define the likelihood ratio:
 \begin{equation}
\mathcal{L}(\mathcal{ \{P\}})=\frac{\tilde{L}(\mathcal{ \{P\}},\tilde{\Omega}_{\mathrm{PT}})}{\tilde{L}(\mathcal{\{P\}},\tilde{\Omega}_{\mathrm{noPT}})}~, 
\label{eq:likeR}
\end{equation} 
where $\tilde{L}$ is the maximized likelihood, $\tilde{\Omega}_{\mathrm{PT}}$ and $\tilde{\Omega}_{\mathrm{noPT}}$ are the optimal template parameters given power spectrum $\{P\}$. In eq.~\ref{eq:likeliPT} and eq.~\ref{eq:likelinoPT}, the frequencies considered for PT-induced signals are from $400~\mathrm{Hz}$ to $950~\mathrm{Hz}$. The eq. \ref{eq:likeR} serves as our \emph{``PT-meter"} to identify the presence of PT. Since the templates $N_{\mathrm{PT}}$ (Eq. \ref{eq:mod2}) and $N_{\mathrm{noPT}}$ (Eq. \ref{eq:mod0}) capture well the main features of the neutrino event rates of the models with and without PT-induced oscillations respectively, as the oscillatory features in the data become more pronounced, the numerator Eq. \ref{eq:likeR} is likely to increase (generally better fit for the PT template), while at the same time the denominator is likely to decrease (poorer fit for the no-PT template), so $\mathcal{L}$ is likely to increase. Vice-versa,  $\mathcal{L}$ will take lower values if the PT-induced oscillatory signals in the data become weaker. The input for the PT-meter is the neutrino power spectrum $\{\mathcal{ P}(k\delta)\}$, and it comes either from predictions of numerical simulations or from future observations of CCSNe neutrinos.

We first use $\{\mathcal{ P}(k\delta)\}$ from two theoretical models to calibrate the $\mathcal{L}$ values of PT-meter in PT and no-PT scenario. The models with and without PT activities are based on the work by \emph{Zha et al.}~\cite{Zha:2021fbi}.  In the following the model with PT activities is denoted as ZOS21-STOS-B145, which is the Zha-O'Connor-Schneider model using the STOS-B145 EoS. And the model without PT activities is denoted as ZOS21-STOS, which is the Zha-O'Connor-Schneider model using the STOS EoS. The STOS EoS is a nucleonic relativistic mean field EoS \cite{2011ApJS..197...20S}. More details about the models with and without PT activities are discussed in section \ref{sec:general}. By considering the shot noise, the neutrino events predicted by the ZOS21-STOS-B145 and the ZOS21-STOS are randomized. In this way, we get the PDF of $\mathcal{L}$ in PT scenario ($\mathrm{Prob}_{\mathrm{PT}}(\mathcal{L})$) and in no-PT scenario ($\mathrm{Prob}_{\mathrm{noPT}}(\mathcal{L})$). 

We previously defined $\mathrm{Prob}_{PT}(\mathcal{L})$ and $\mathrm{Prob}_{\mathrm{noPT}}(\mathcal{L})$. In the future, given $\{\mathcal{ P}(k\delta)\}$ from observations of CCSNe neutrinos we might have  $\mathcal{L}_{\mathrm{obs}}$, which is the observed PT-meter value. We further define the threshold of PT-meter, which is $\mathcal{L}_{thr}$. The PT exists if $\mathcal{L}_{\mathrm{obs}}>\mathcal{L}_{\mathrm{thr}}$ and doesn't exist if $\mathcal{L}_{\mathrm{obs}}<\mathcal{L}_{\mathrm{thr}}$. The probability of correctly claiming the PT existence is:
\begin{equation}\label{eq:pd}
 P_D=\int^\infty_{\mathcal{L}_{\mathrm{thr}}}d\mathcal{L}~\mathrm{Prob}_{\mathrm{PT}}(\mathcal{L}),   
\end{equation}
 and the false alarm rate is:
 \begin{equation}\label{eq:pfa}
    P_{FA}=\int^\infty_{\mathcal{L}_{thr}}d\mathcal{L}~\mathrm{Prob}_{\mathrm{noPT}}(\mathcal{L}). 
 \end{equation}

Here we introduce the PDF of the PT template parameters $\tilde{\Omega}_{PT}$. Note when calculating $\mathcal{L}$, the relationship between $\{\mathcal{ P}(k\delta)\}$ and $\tilde{\Omega}_{PT}$ is a one-to-one relationship. We determine the PDF of $\tilde{\Omega}_{PT}$  by taking many realizations and therefore sampling the distribution from the shot noise. In our Monte Carlo simulations, we found this PDF based on the ZOS21-STOS-B145 and the ZOS21-STOS at various CCSNe distances. The mean of $\tilde{\Omega}_{PT}$ relate to the features of PT oscillations (such as oscillation amplitude and frequencies). And the variation of $\tilde{\Omega}_{PT}$ relates to the uncertainties associated with our determination of the PT oscillation features due to the shot noise inherent in any observation.

\section{result}
\label{sec:result}
\subsection{Time series and power spectrum}
\label{sub:discretized}

In this section we present the time series and power spectrum of the neutrino signals predicted by the ZOS21-STOS-B145 and the ZOS21-STOS models in IceCube and Hyper-K. 

\begin{figure*}[ht]
	\centering

	\includegraphics[width=0.45\textwidth]{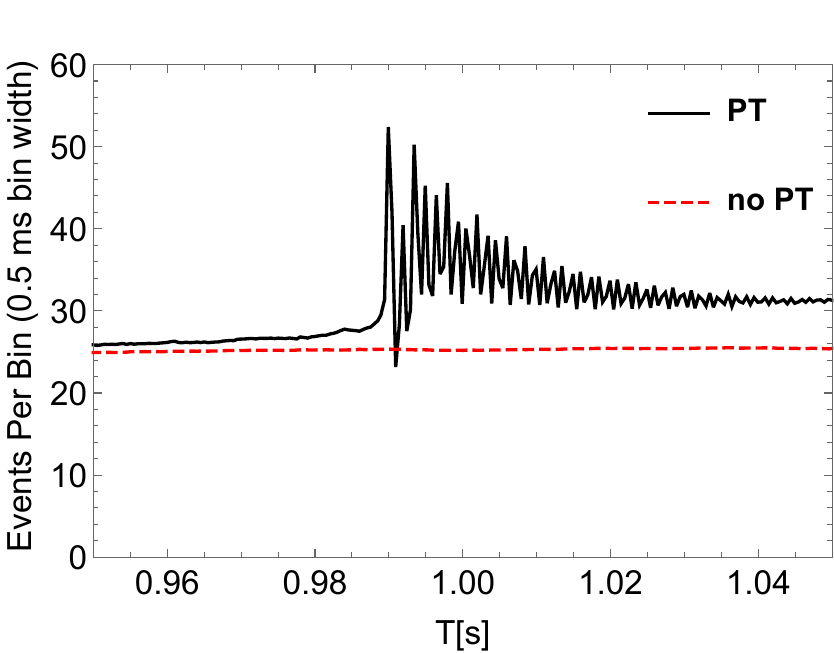}
	\includegraphics[width=0.45\textwidth]{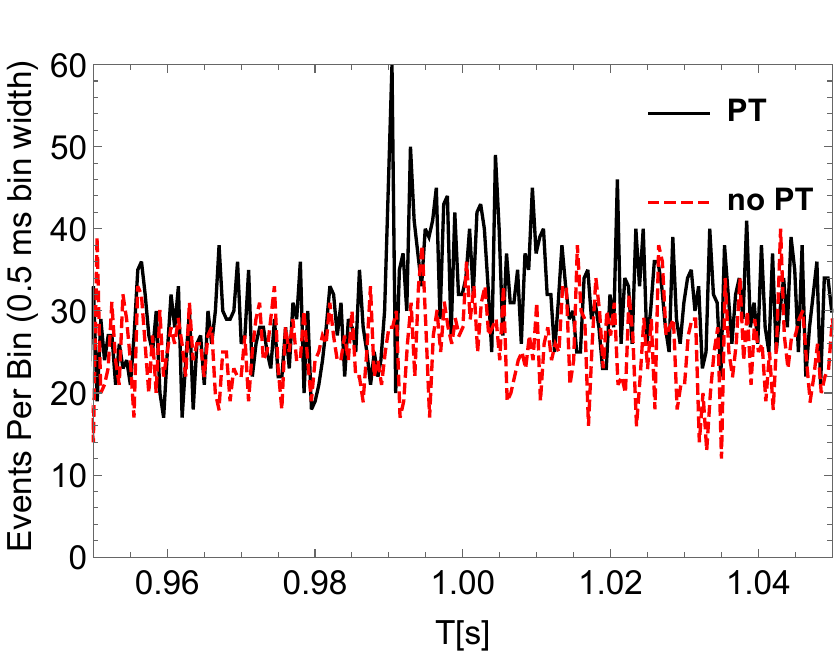}
	\includegraphics[width=0.45\textwidth]{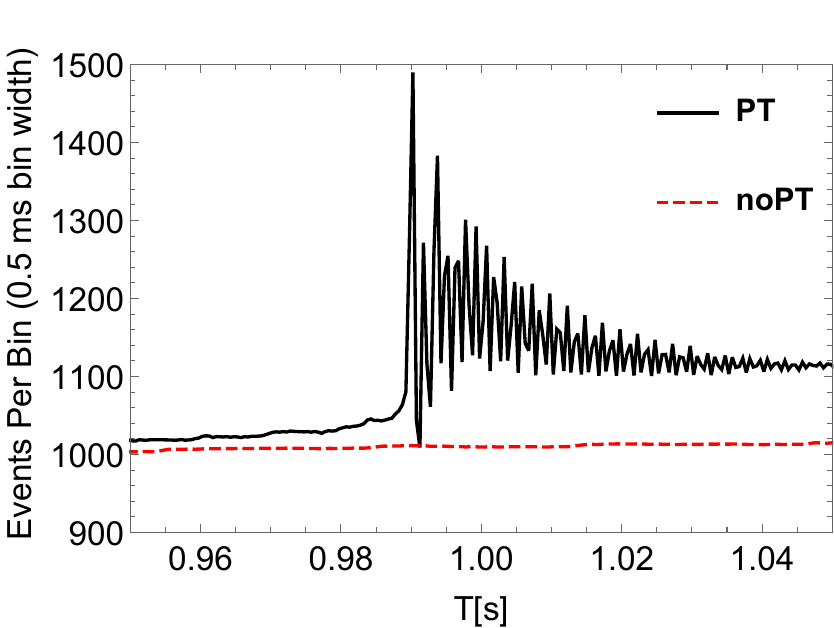}
	\includegraphics[width=0.45\textwidth]{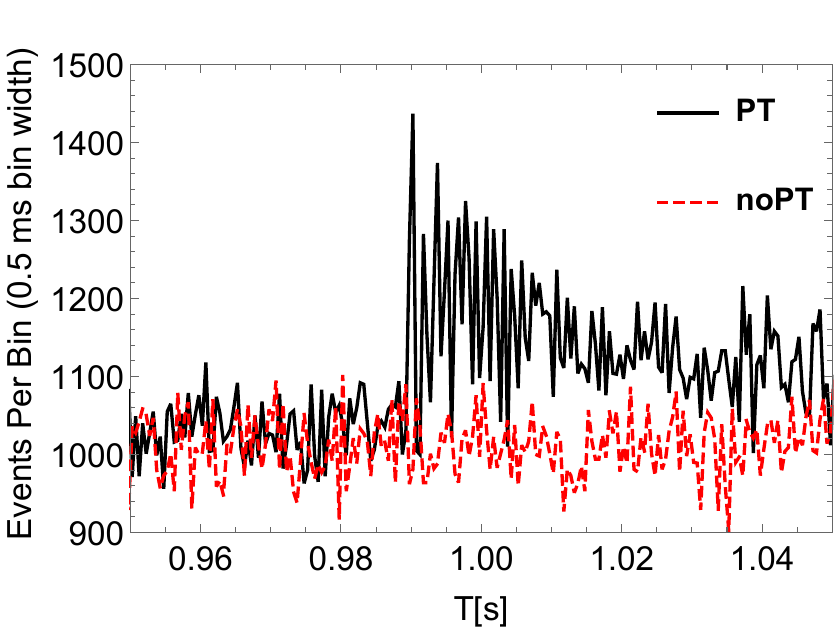}
	
	\caption{Neutrino event rate at \hk\ (upper panels) and IceCube (lower panels), with (right) and without (left) the shot noise at SNe distance $D=10$\,kpc. }
	\label{fig:hypiceT}
\end{figure*}

\begin{figure*}
	\centering

	\includegraphics[width=0.45\textwidth]{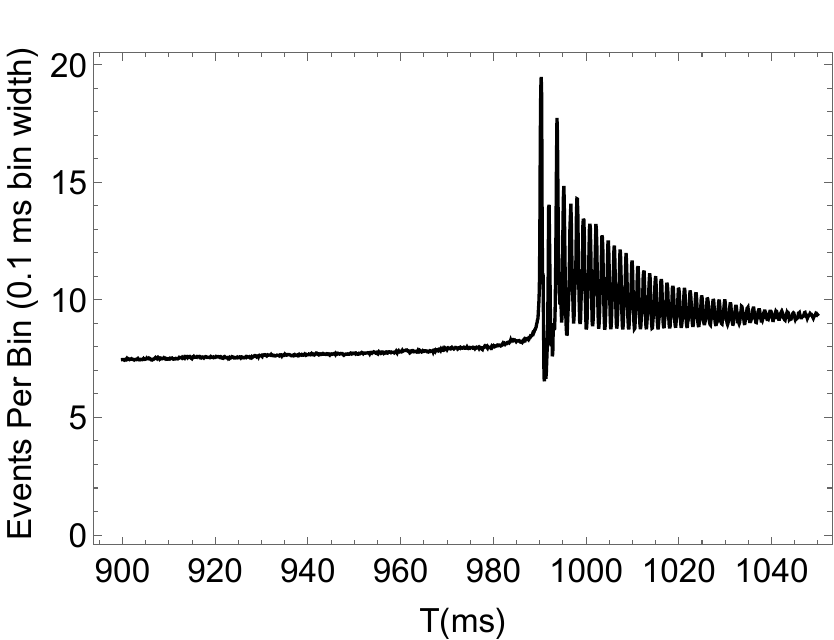}
	\includegraphics[width=0.45\textwidth]{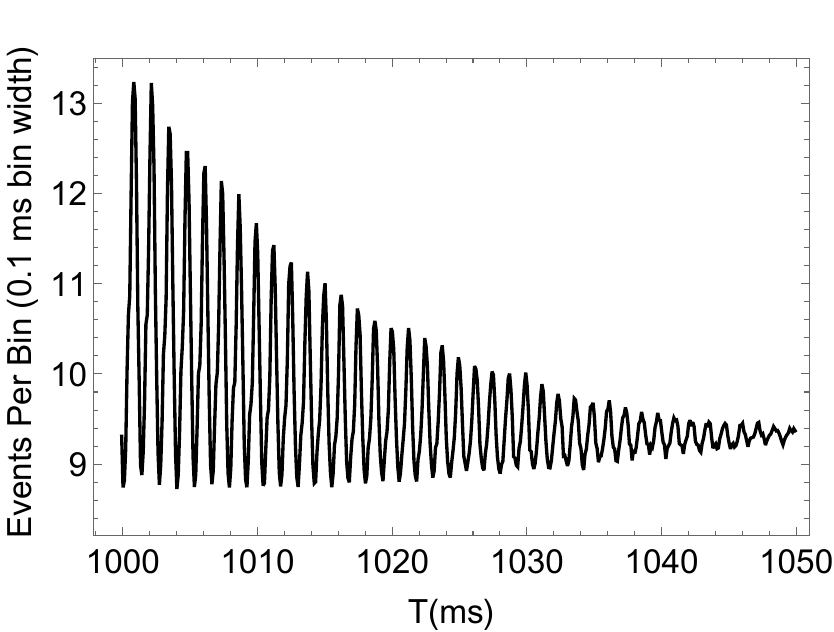}
	\includegraphics[width=0.45\textwidth]{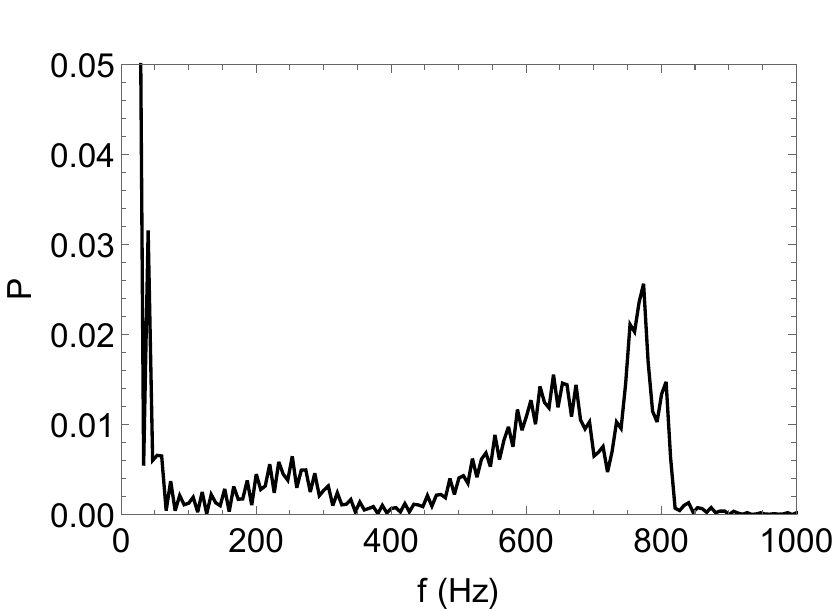}
	\includegraphics[width=0.45\textwidth]{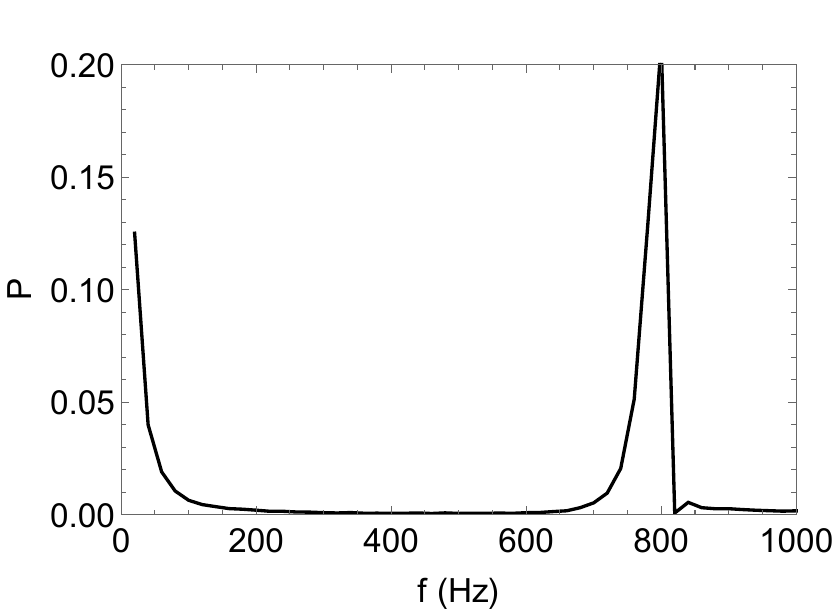}
	\caption{Neutrino event rate (upper 2 panels) and its power spectrum (lower 2 panels) based on ZOS21-STOS-B145 at \hk\,from $900$ to $1050$ ms after bounce (left) and from $1000$ to $1050$ ms after bounce (right), for distance $D=10$ kpc. }
	\label{fig:hypTandP}
\end{figure*}

Let us first discuss qualitatively the features of PT-induced neutrino signals in the time and frequency domain. As shown in Fig.~\ref{fig:hypiceT} and Fig.~\ref{fig:hypTandP}, the PT-induced neutrino signal in this model manifests itself by presenting a nearly monochromatic oscillation of the number of neutrino events from approximately $1000~\mathrm{ms}$ to $1040~\mathrm{ms}$ postbounce in the time domain and correspondingly has a power peak at around $800$\,Hz in the frequency domain. Additionally, the PT activity itself gives a strong burst at the beginning of the PT, around $\sim 990~\mathrm{ms}$ in the time domain. Before stabilizing, this burst contains the first several oscillation cycles from $\sim 990~\mathrm{ms}$ to $\sim 1000~\mathrm{ms}$ and forms a peak at around $650~\mathrm{Hz}$ in frequency domain, which is slightly different from the main PT-induced peak at $\sim 800~\mathrm{Hz}$. Note that the PT-induced oscillation observed in detectors is unavoidably smeared by shot noise. In Fig. \ref{fig:hypiceT}, we compared the neutrino signals with and without shot noise in time domain. Since the relative uncertainty due to shot noise is $\propto \sqrt{N}/N \propto D$, as the SNe distance $D$ increases the PT-induced oscillation become more difficult to be distinguished from the observed neutrino events in the time domain. The existence and the feature of PT-induced oscillation may be easier to quantitatively analyze in frequency domain.  

We then discuss the qualitative features of the neutrino signal in both the time and the frequency domain with several sets of $t_{\mathrm{ST}}$ and $\tau$ to be used when searching for the PT-induced signals. For example, as shown in the upper left panel of Fig. \ref{fig:hypTandP}, a time series with $\tau\approx150~\mathrm{ms}$ and $t_{\mathrm{ST}}\approx900~\mathrm{ms}$ includes the neutrino signal before and during the PT-induced oscillations. Correspondingly the burst indicating the start of PT-induced oscillations in time domain can clearly be observed. In the lower left panel of Fig. \ref{fig:hypTandP} we show the power spectrum of this neutrino signal. The very first several oscillation cycles with highest amplitude from $990~\mathrm{ms}$ to $1000~\mathrm{ms}$ gives a peak at $\sim650~\mathrm{Hz}$ in frequency domain as shown in the lower left panel of Fig.~\ref{fig:hypTandP}. As will be shown in section \ref{sec:result}, in the PT-meter analysis the lower-frequency peak at $\sim ~650 ~\mathrm{Hz}$ serves as an additional signal indicating the start of PT-oscillations, and helps to determine the existence as well as the starting time of PT-induced neutrino oscillations. In the right panels of Fig. \ref{fig:hypTandP}, we show the neutrino signal from $1000~\mathrm{ms}$ to $1050~\mathrm{ms}$ (top) and its power spectrum (bottom). Only stable PT-induced oscillation are included in this time range, and only a main peak at $\sim 800~\mathrm{Hz}$ is observed in the power spectrum. More details will be discussed in the next section.

\subsection{Are there PT-induced quasiperiodic neutrino signals?}
\label{sub:results}
\begin{figure*}
	\centering

	\includegraphics[width=0.45\textwidth]{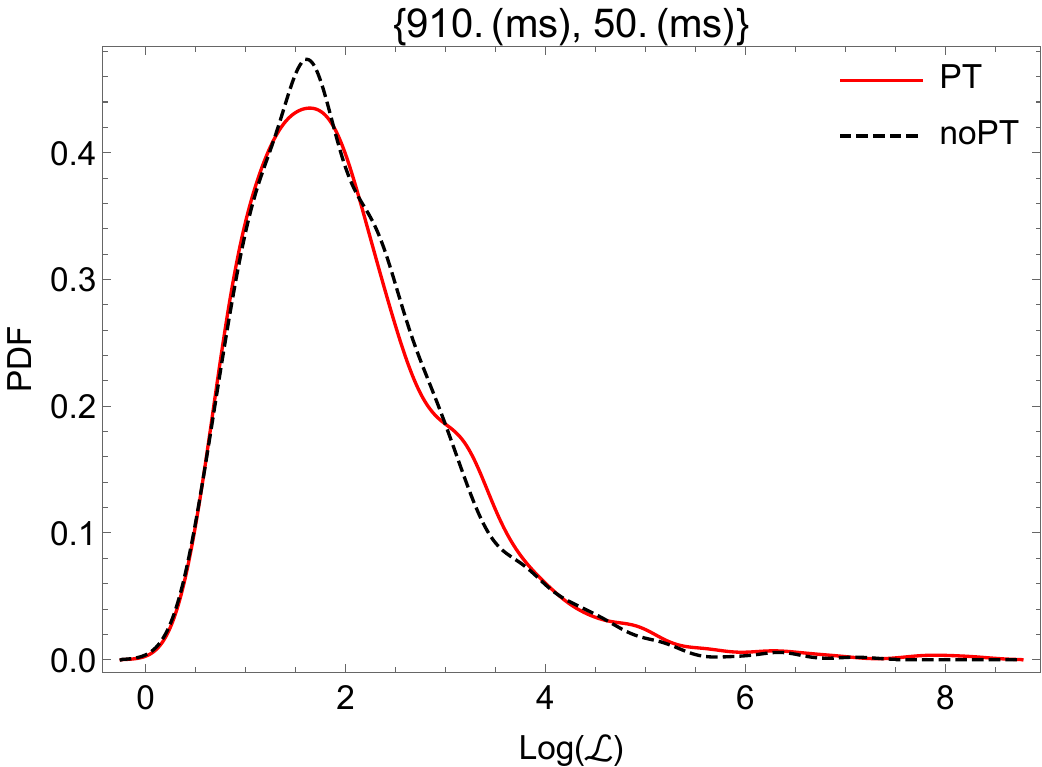}
	\includegraphics[width=0.45\textwidth]{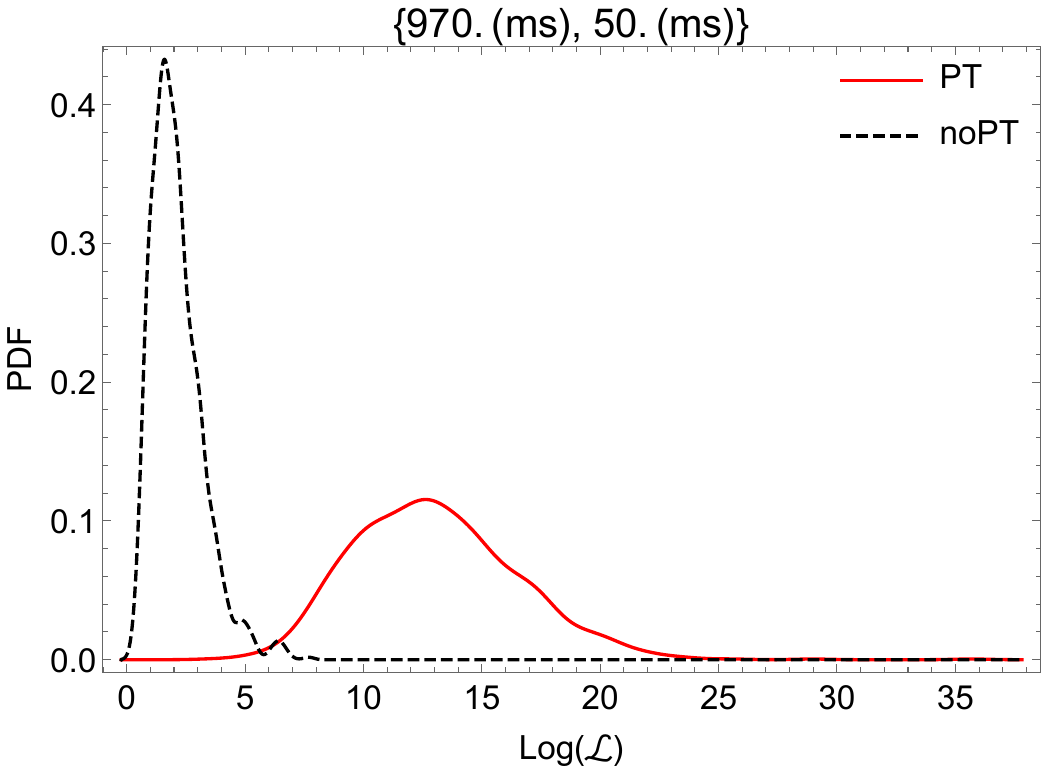}
	\includegraphics[width=0.45\textwidth]{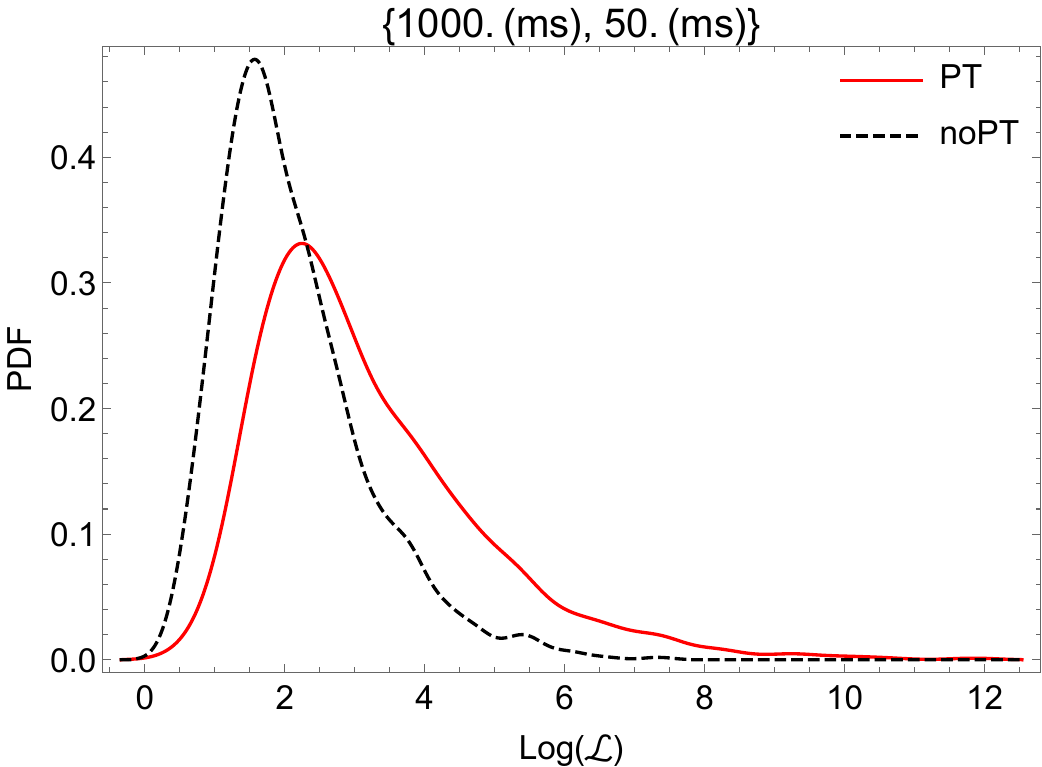}
	\includegraphics[width=0.45\textwidth]{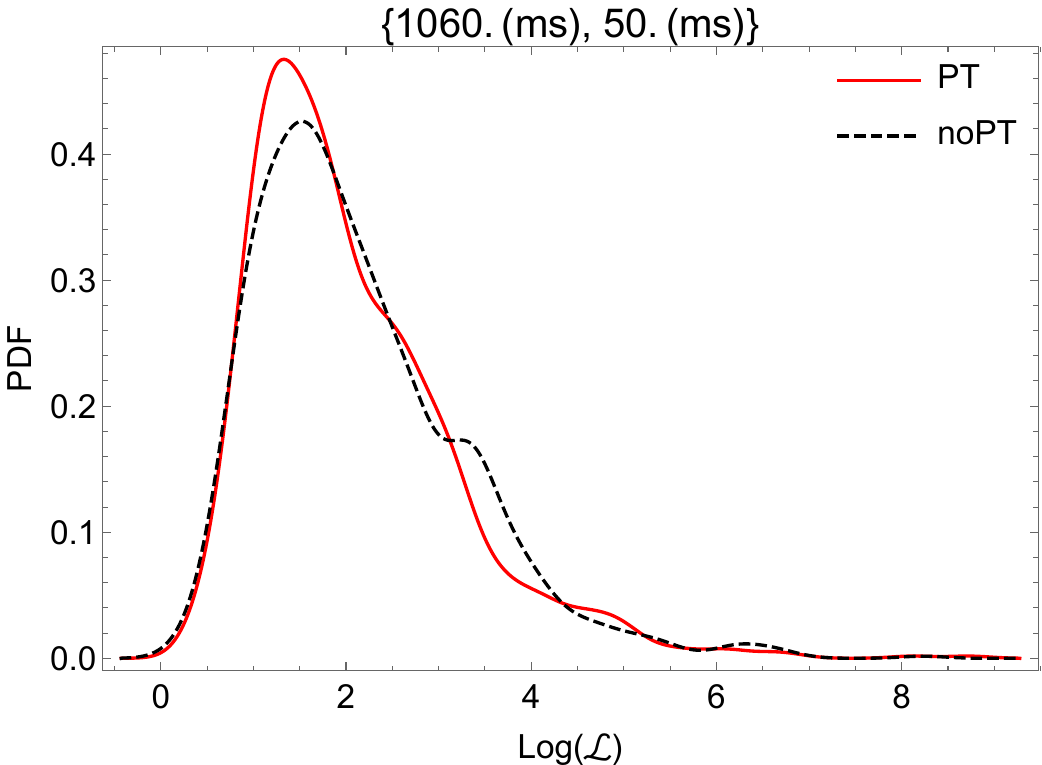}
	
	\caption{Probability density distribution of $\mathcal{L}$ in Hyper-K at 5 kpc. The PDF of $\mathcal{L}$ is estimated based on the ZOS21-STOS-B145 and ZOS21-STOS models, in both PT (Red) and no-PT (black dashed) scenario. The upper left, upper right, lower left and lower right panel corresponds to the PDF of $\mathcal{L}s$ before the PT-induced oscillation, during the PT-induced oscillation transition, during the PT-induced oscillation and after the PT-induced oscillation. The starting time and the duration of analyzed time series  are shown at the top of each panel.}
	\label{fig:likeRPDF}
\end{figure*}

\begin{figure*}[htp]
	\centering
	\includegraphics[width=0.45\textwidth]{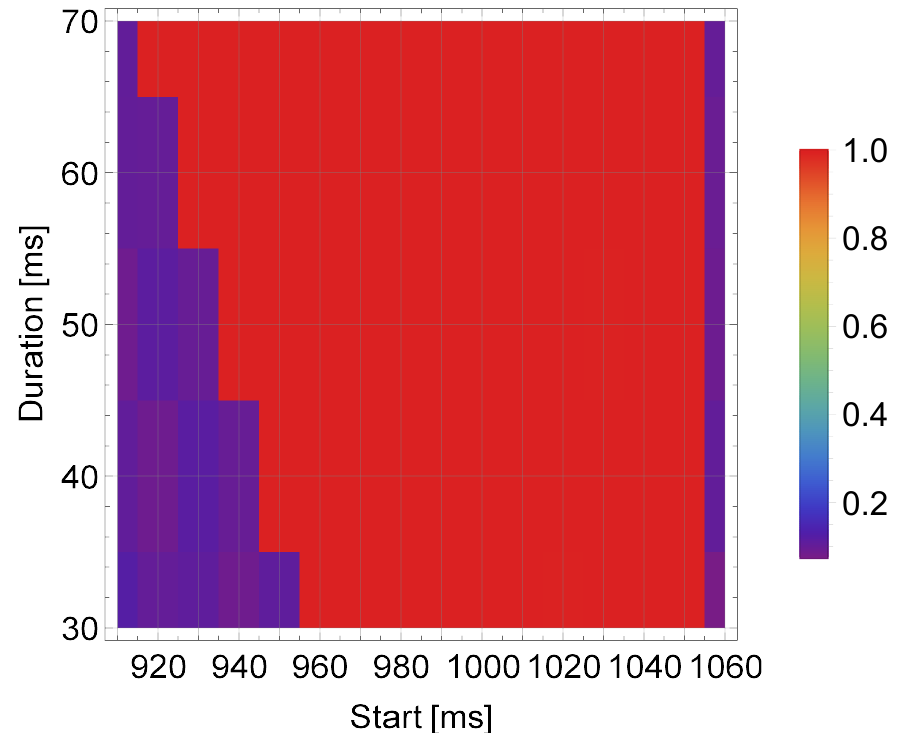}
	\includegraphics[width=0.45\textwidth]{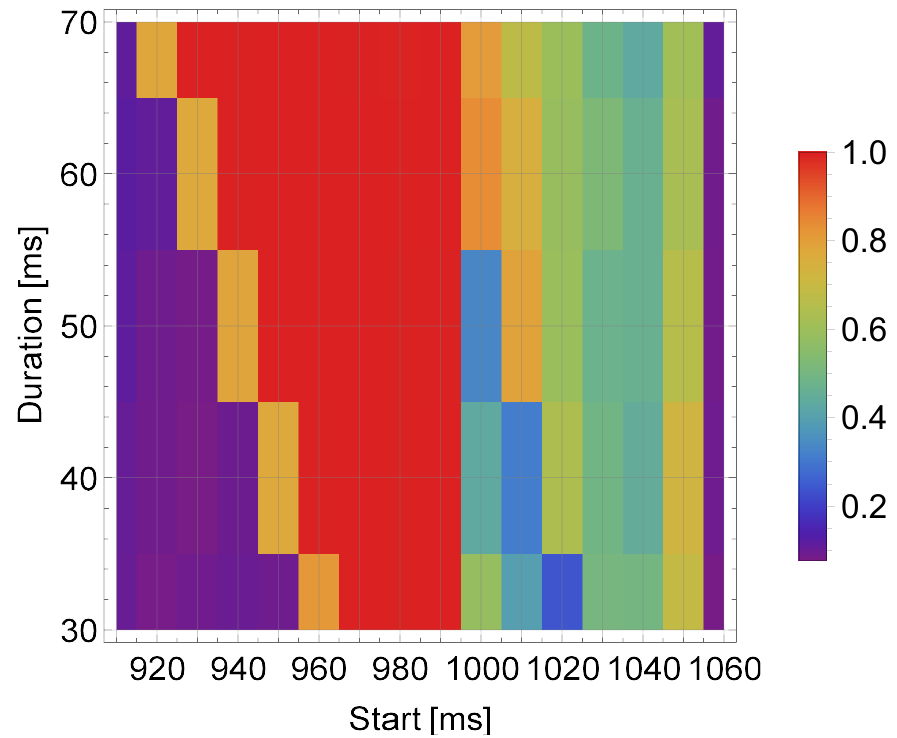}
	\includegraphics[width=0.45\textwidth]{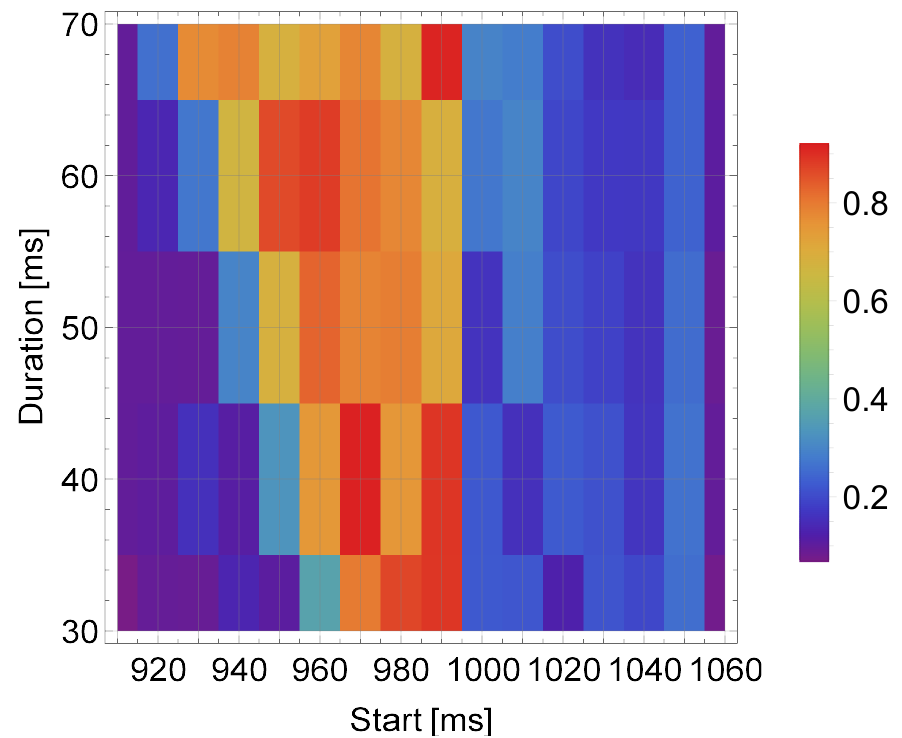}

	\caption{Probability of detection at false alarm rate being $10\%$, at 1(upper left),5 (upper right) and 10 (lower) kpc in Hyper-K. The vertical and horizontal axis are duration and starting time of analyzed neutrino time series respectively.} 
	\label{fig:PDmap}
\end{figure*}

\begin{figure*}[htp]
	\centering
	\includegraphics[width=0.45\textwidth]{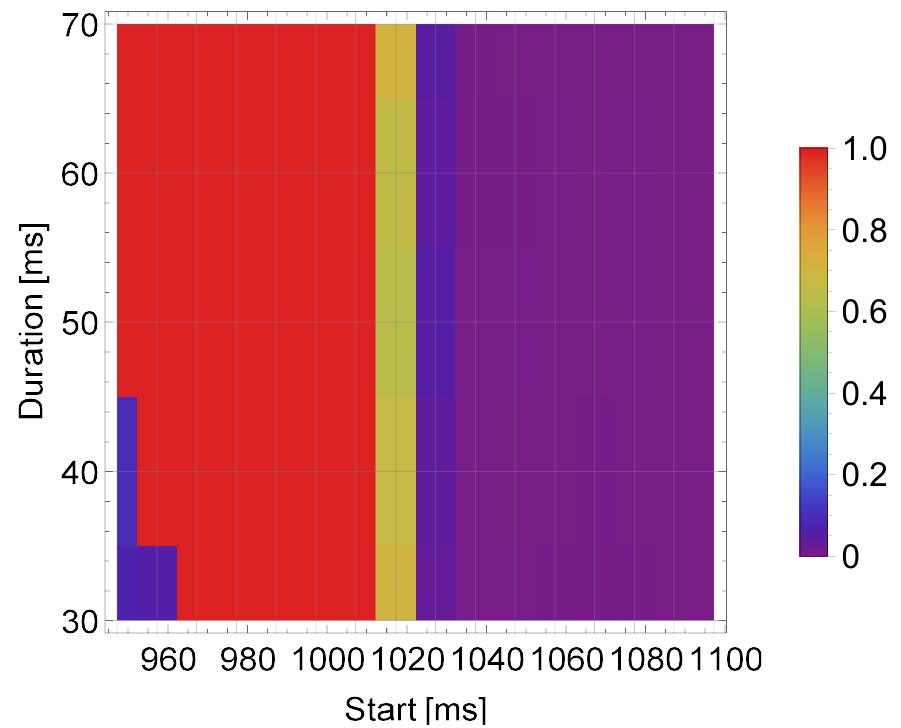}
	\includegraphics[width=0.45\textwidth]{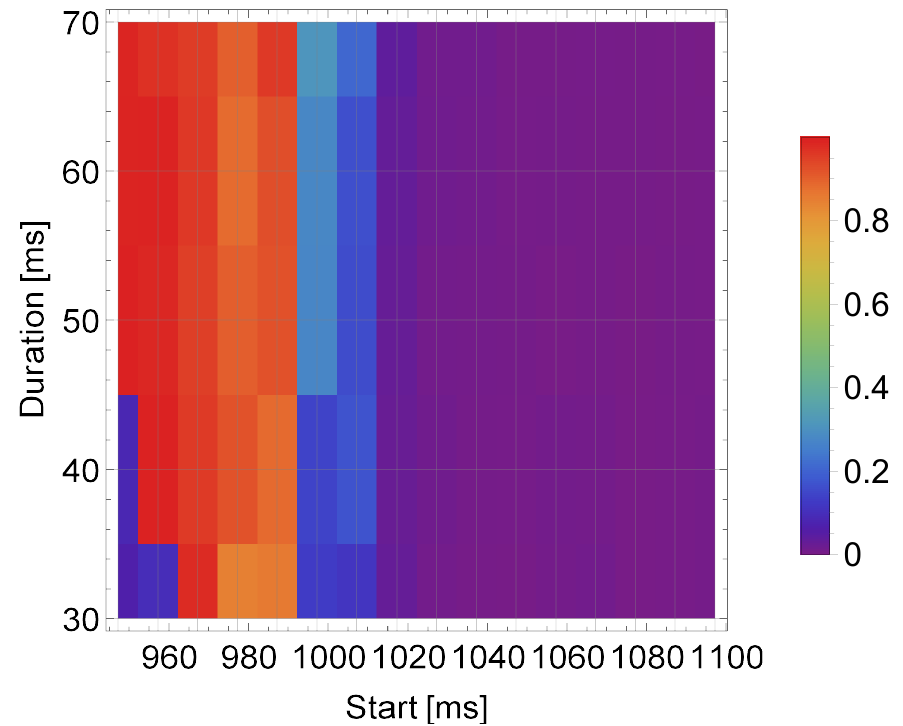}

	\caption{Probability of detection at false alarm rate being $10\%$, at 5(left), and 10 (right) kpc in IceCube. The vertical and horizontal axis are duration and starting time of analyzed neutrino time series respectively. } 
	\label{fig:PDmapIce}
\end{figure*}

As discussed in section \ref{sec:maxlik}, the duration and starting time of the analyzed time series of neutrino events can be tuned by changing $t_{\mathrm{ST}}$ and $\tau$ when calculating $\mathcal{L}$. In each time series the PDFs of $\mathcal{L}$ based on the PT and no-PT numerical simulations are generated. In Fig. \ref{fig:likeRPDF}, we present the PDF of $\mathcal{L}$ in two different scenarios (PT and no-PT) for four representative time series (noted at the top of each panel) for a source at 5\,kpc. The upper left and the lower right panel of Fig. \ref{fig:likeRPDF} represent PDFs of $\mathcal{L}$ for time series before and after the PT-induced oscillation. The PDFs of $\mathcal{L}$ based on the ZOS21-STOS-B145 and ZOS21-STOS in these two panels are not distinguishable. The upper right panel of Fig. \ref{fig:likeRPDF} represents PDFs of $\mathcal{L}$ for time series including the burst. Due to the large amplitude in the first several $\mathrm{ms}$ of PT-induced oscillation, the PDFs of $\mathcal{L}$ based on the ZOS21-STOS-B145 and ZOS21-STOS models can be easily distinguished here. The lower left panel of Fig. \ref{fig:likeRPDF} represents PDFs of $\mathcal{L}$ for time series during the stable PT-induced oscillation. Although the PDFs of $\mathcal{L}$ based on PT and no-PT models are different, one may not distinguish them with high statistical confidence in this time series based on the neutrino events from a CCSNe at 5 kpc in Hyper-K due to the damping of PT-induced oscillation. Indeed, as shown in the upper left panel of Fig. \ref{fig:PDmap}, the $P_D$ with $P_{FA}=10\%$ is only $\approx 30\%$ in this specific time series. The $P_D$ is larger in time series with similar starting time but longer duration. Note the discrete frequency resolution $\delta=1/\tau$, and the likelihood $L(\{\mathcal{P}\}, \Omega_{PT})$ is a product of probabilities at different discrete frequencies. So, having longer duration and thus finer discrete frequency resolution may increase the $P_D$ in time series with similar starting time.  Given PDFs of $\mathcal{L}$ in two scenarios, a threshold of $\mathcal{L}_{thr}$ can be determined in various time series in accordance with designated $P_{FA}$, which is defined in eq.~\ref{eq:pfa}. Given $\mathcal{L}_{thr}$, $P_D$ defined in eq.~\ref{eq:pd} can be found in various time series. In Fig.~\ref{fig:PDmap} and Fig.~\ref{fig:PDmapIce}, we present a map of $P_D$ values at false alarm rate $P_{FA}=10\%$ in time series with different duration and starting time.

Let us now discuss the $P_D$ map of Hyper-K at 5 kpc, for illustration purposes. The analyzed neutrino signals can be mainly classified into 3 types: (1) the analyzed time series ends before or starts after the PT-induced oscillation; (2) the analyzed time series starts before the PT-induced oscillation and ends during the oscillation; (3) the analyzed time series starts during the PT-induced oscillation and ends during or after the oscillation. Correspondingly, based on the magnitude of $P_D$, the map for Hyper-K at 5 kpc can be roughly divided into 3 regions: (1) the region where $P_D\approx0$; (2) region where $0.9<P_D<1$ and (3) region where $0.4<P_{D}<0.9$. We see that when the analyzed time series includes the burst, it has highest $P_D$ and thus highest ability of distinguishing a PT from a no-PT scenario. This can be understood easily since the oscillation amplitude at the beginning is large. The region corresponding to the time series that start during the oscillation has the second highest $P_D$, while the region corresponding to time series ending before or starting after the oscillation has $P_D\approx0$ and is obviously lower than the $P_D$ values of the former 2 regions. Similar behavior of $P_D$ distribution can be observed in maps for both Hyper-K and IceCube at various SNe distances. 

We now discuss the application of these $P_D$ maps for the future detection of the PT-induced neutrino oscillations. The PD maps (Fig. \ref{fig:PDmap} and Fig. \ref{fig:PDmapIce}) help one to find a time series with specific $t_{\mathrm{ST}}$ and $\tau$ where the PT-induced oscillations can be distinguished from no-PT signals with high $P_D$. For example, in the upper right panel of Fig. \ref{fig:PDmap}, the best candidates for testing PT-induced signals are the time series including the burst at around $990~\mathrm{ms}$ postbounce. 

One may notice the distribution of $P_D$ values is unavoidably model-dependent, since the PDFs of $\mathcal{L}$ in both PT and no-PT scenarios are (and have to be) based on theoretical predictions of PT/no-PT signals from numerical simulations. We leave further discussion on the model dependency of the PT distinction to section \ref{sec:disc}. 

\subsection{Parameter estimation}
\label{sec:parameter}

In this section we discuss the parameter estimation using PT-meter. Following the method described in section \ref{sec:maxlik}, we obtained the PDFs of optimal template parameters $\tilde{\Omega}_{\mathrm{PT}}=\{\tilde{a},\tilde{f}_{\rm PT}\}$.

\begin{figure*}[htp]
	\centering
	\includegraphics[width=0.45\textwidth]{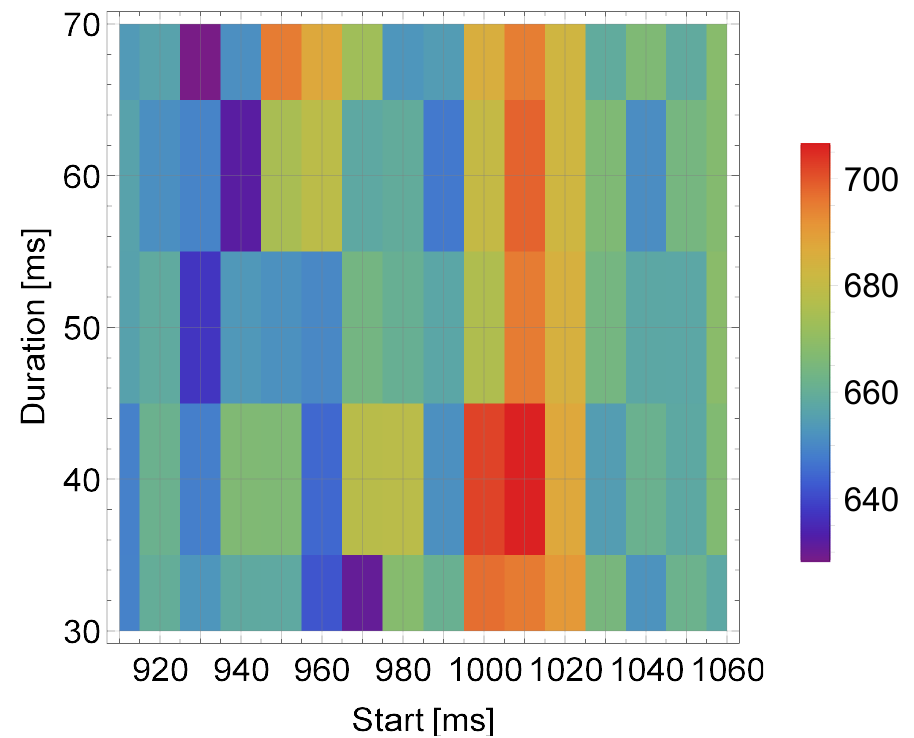}
	\includegraphics[width=0.45\textwidth]{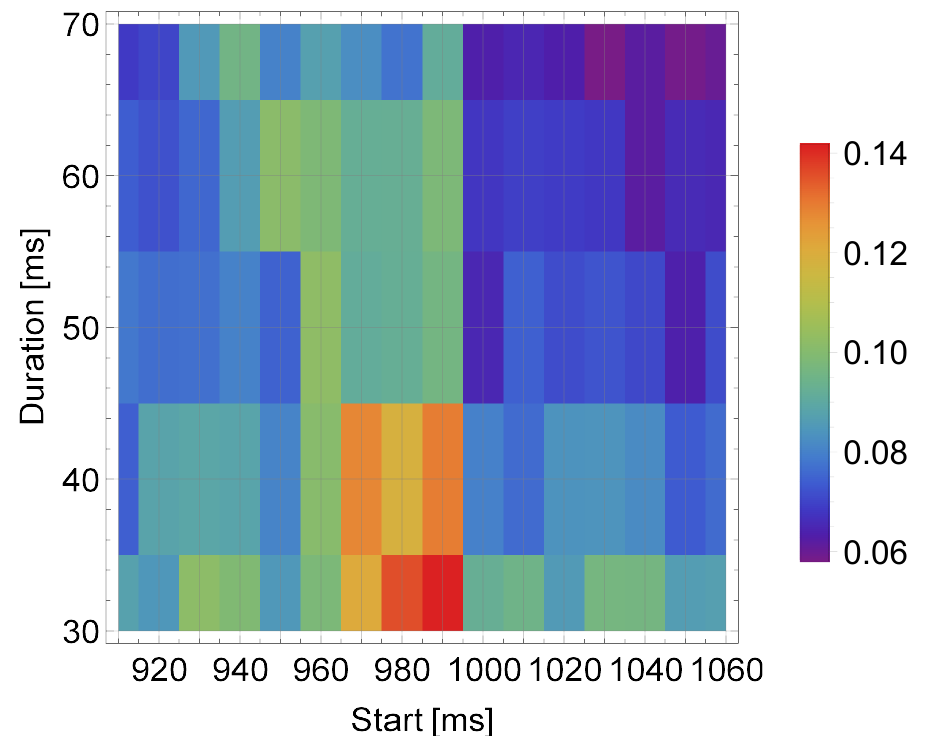}

	\includegraphics[width=0.45\textwidth]{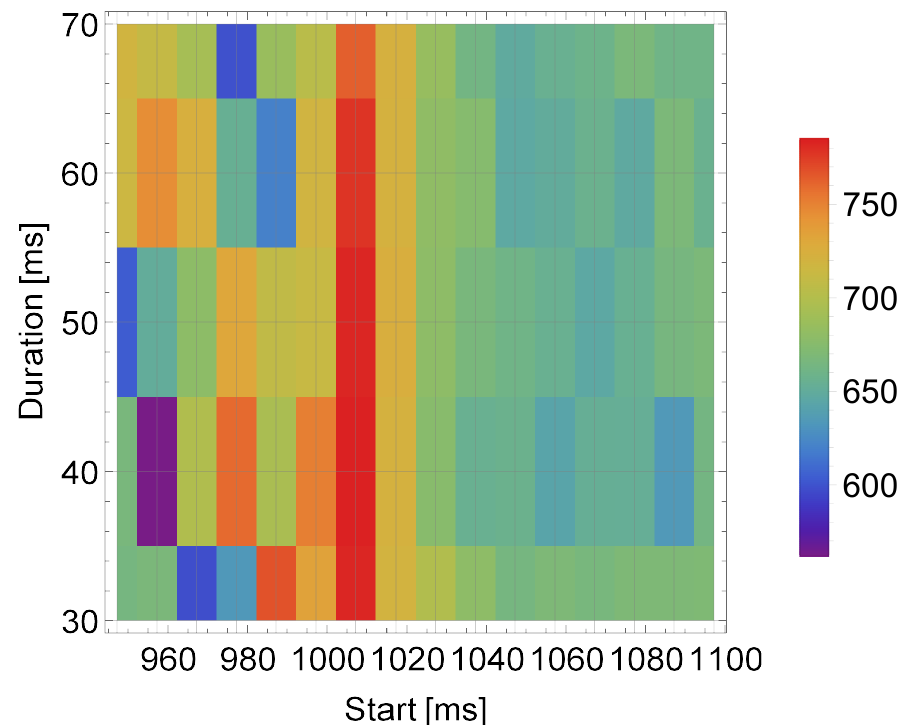}
	\includegraphics[width=0.45\textwidth]{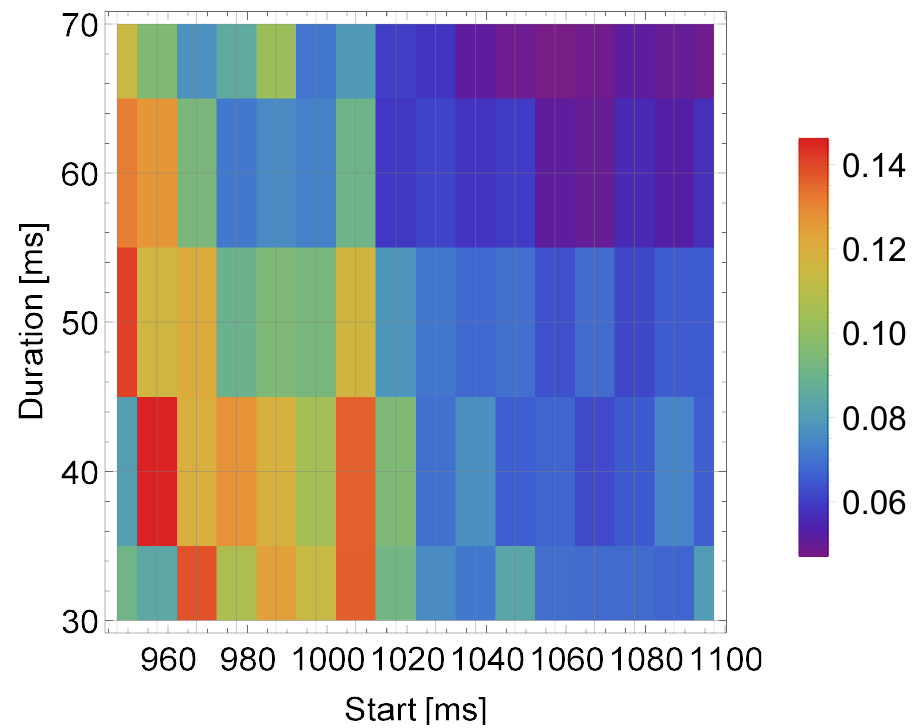}

	\caption{Averaged frequencies $\tilde{f}_{\rm PT}$ (left) and amplitude $a$ (right) in various time series, at 10 kpc in Hyper-K (upper) and IceCube (lower). The vertical and horizontal axis are duration and starting time of analyzed neutrino time series respectively. } 
	\label{fig:faPT10kpcmap}
\end{figure*}

\begin{figure*}[htp]
	\centering
	\includegraphics[width=0.45\textwidth]{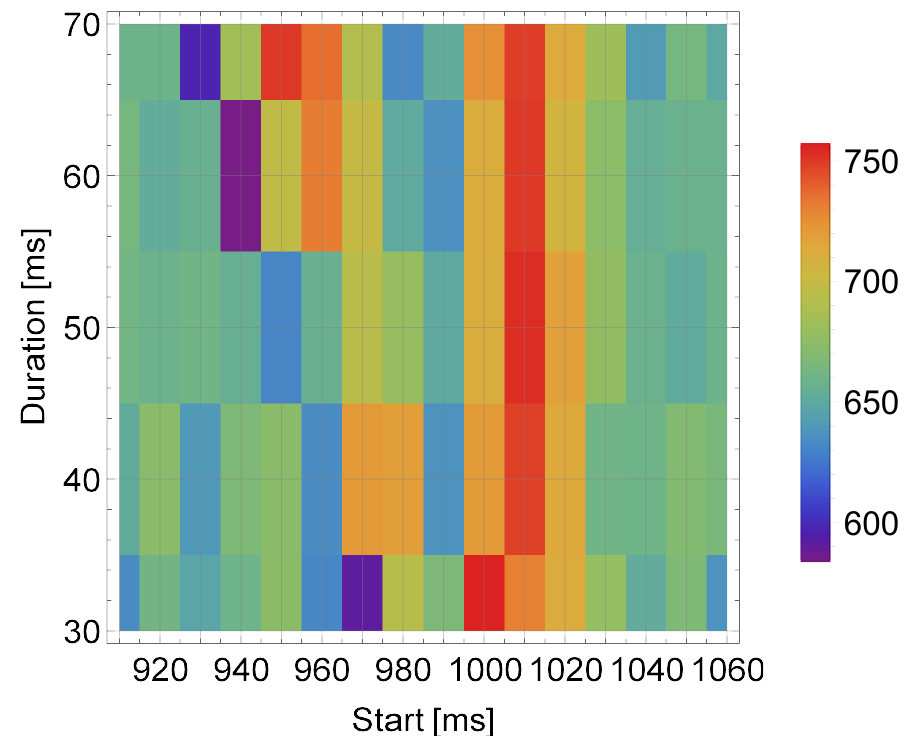}
	\includegraphics[width=0.45\textwidth]{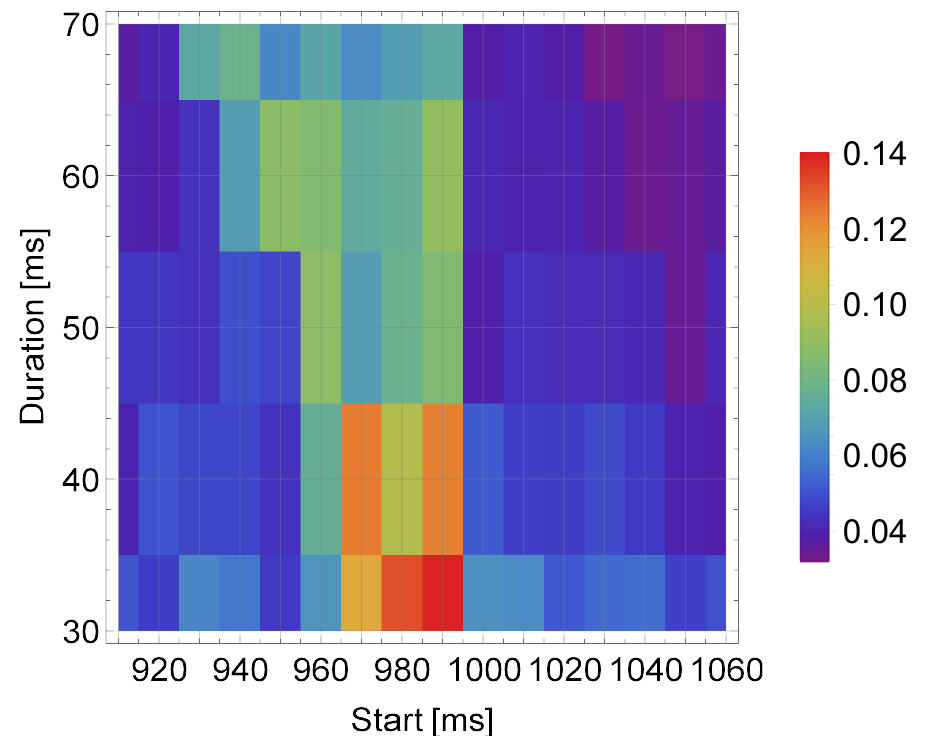}
	
	\includegraphics[width=0.45\textwidth]{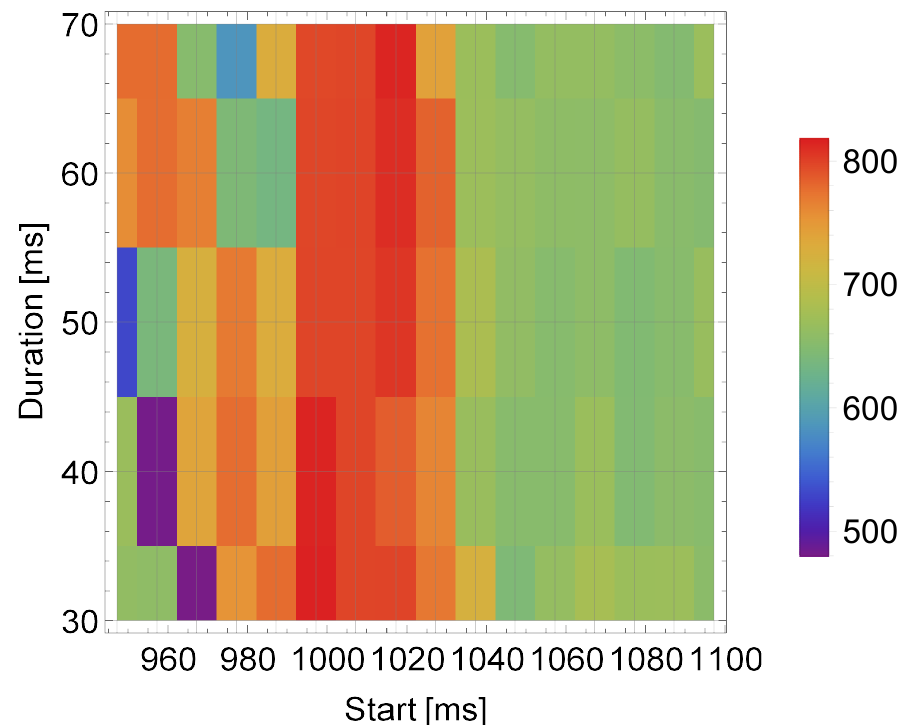}
	\includegraphics[width=0.45\textwidth]{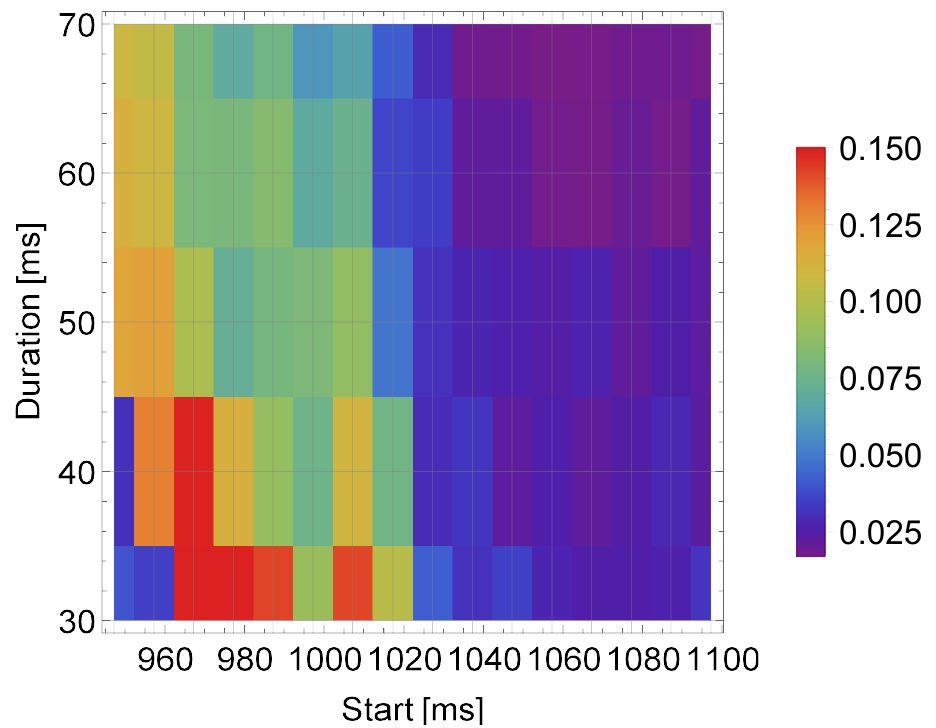}

	\caption{Averaged frequencies $\tilde{f}_{\rm PT}$ (left) and amplitude $a$ (right) in various time series, at 5 kpc in Hyper-K (upper) and IceCube (lower). The vertical and horizontal axis are duration and starting time of analyzed neutrino time series respectively. } 
	\label{fig:faPT5kpcmap}
\end{figure*}

\begin{figure*}[htbp]
	\centering
	\includegraphics[width=0.45\textwidth]{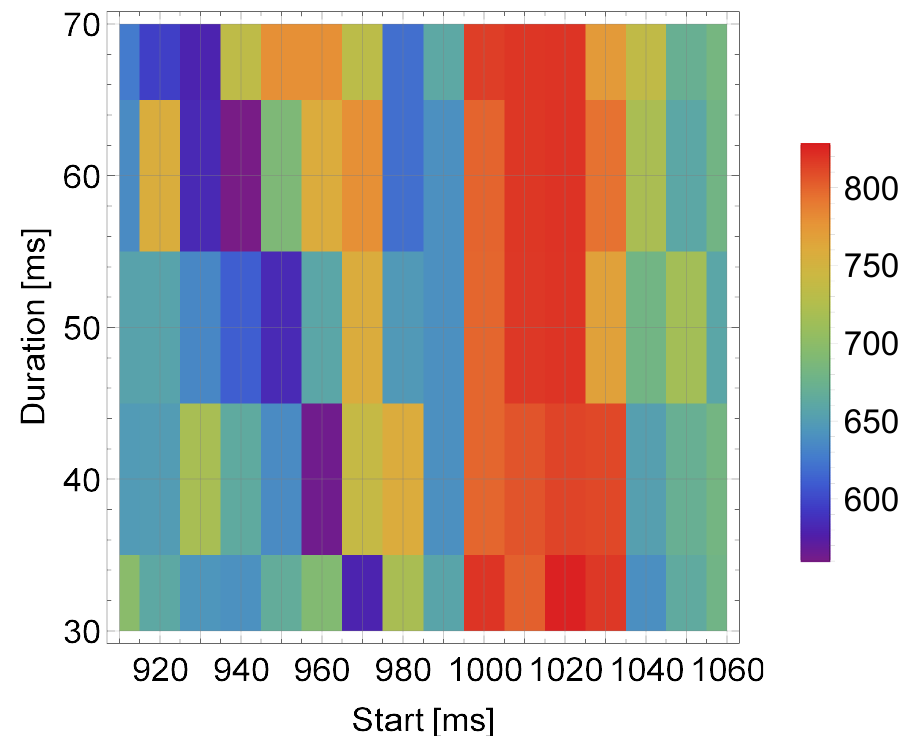}
	\includegraphics[width=0.45\textwidth]{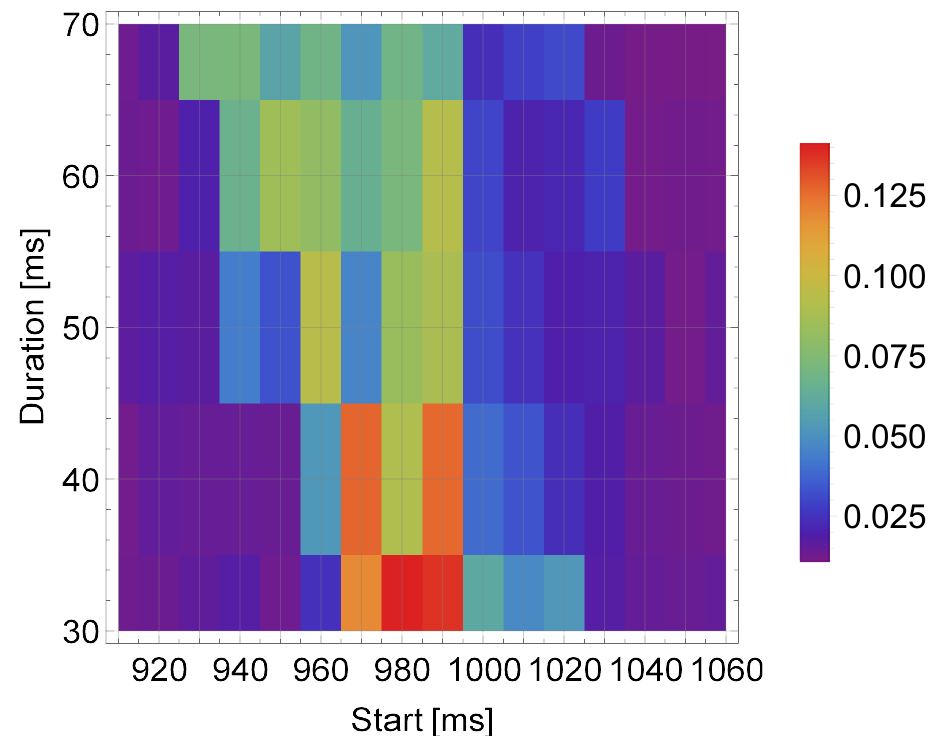}
	
	\caption{Averaged frequencies $\tilde{f}_{\rm PT}$ (left) and amplitude $a$ (right) in various time series, at 1 kpc in Hyper-K. The vertical and horizontal axis are duration and starting time of analyzed neutrino time series respectively. } 
	\label{fig:faPT1kpcmap}
\end{figure*}

We then estimate the mean and the standard deviations of the nominal oscillation frequency $\tilde{f}_{\mathrm{PT}}$ and amplitude $\tilde{a}$, in IceCube at 10 and 5\,kpc and in Hyper-K at 10, 5, and 1\,kpc. In Fig.~\ref{fig:faPT10kpcmap}, Fig.~\ref{fig:faPT5kpcmap} and Fig.~\ref{fig:faPT1kpcmap}, for 10, 5, and 1\,kpc, respectively, we show the averaged $\tilde{f}_{\mathrm{PT}}$ in various neutrino time series. Additionally, in Tab.~\ref{tab:hyppara} and Tab.~\ref{tab:hyppara1kpc}, the mean and the standard deviations of $\tilde{f}_{\mathrm{PT}}$ and $\tilde{a}$ for several selected time series in Hyper-K at 1\,kpc and 10 kpc,  respectively. 

For illustration purposes, we discuss the maps for Hyper-K at 1\,kpc in Fig. \ref{fig:faPT1kpcmap}. We see that the time series with starting times of $1000~\mathrm{ms}<t_{\mathrm{ST}}<1030~\mathrm{ms}$ and durations of $30~\mathrm{ms}<t_{\mathrm{dur}}<70~\mathrm{ms}$ have approximately the same optimal frequencies at $\approx 800~\mathrm{Hz}$, forming a monochromatic \emph{"cluster"} in the $\tilde{f}_{\mathrm{PT}}$ map. Such a \emph{"cluster"} indicates that this is the region including only PT-induced signals, because of the monochromatic feature of  PT-induced oscillation. Additionally, we see that $\tilde{a}$ is obviously higher in the region with $t_{\mathrm{ST}}\approx 980~\mathrm{ms}$ and $t_{\mathrm{dur}}\approx30~\mathrm{ms}$.  This is the region that includes the burst in between the no-PT and PT time series. Indeed, it is the oscillation in the first several $\mathrm{ms}$ of PT in the burst that induces large $\tilde{a}$ in this region. Additionally, the $\tilde{a}$ of time series starting at identical $t_{\mathrm{ST}}$ monotonically decreases as $t_{\mathrm{dur}}$ increases, because of the damping of PT-induced oscillatory signals observed in the ZOS21-STOS-B145 model. Note for $\tilde{f}_{\mathrm{PT}}$ evaluated at higher distances, its mean may deviate from the theoretically predicted PT frequencies. Indeed, at 10 kpc, as shown in Tab. \ref{tab:hyppara}, the mean of $\tilde{f}_{\mathrm{PT}}$ corresponding to time series with $t_{\mathrm{ST}}=1010~\mathrm{ms}$ and $t_{\mathrm{dur}}=30~\mathrm{ms}$ is approximately 700 Hz, which is lower than the theoretically predicted PT frequency $\sim 800 \mathrm{Hz}$ in the ZOS21-STOS-B145 model with PT. At 10 kpc, the shot noise induces a PDF of $\tilde{f}_{\mathrm{PT}}$ that spread over a wide range of values. Thus, the impact of finite range of chosen available frequency (which is 400-950 Hz in this work) cannot be ignored on the mean of $\tilde{f}_{\mathrm{PT}}$. Such a \emph{"finite size effect"} is not surprising in discrete signal processing. Indeed, the Nyquist frequency $f_{\mathrm{nyq}}=1/2\Delta$ defines the upper bound of the chosen available frequency. In this work with $\Delta=0.5 ~\mathrm{ms}$, the $f_{\mathrm{nyq}}=1000~\mathrm{Hz}$. To estimate the possible impact of finite range of chosen available frequencies, we calculated the standard deviation, $\delta f$. If the range of $\tilde{f}_{\mathrm{PT}}\pm 3\delta f$ is well within the available frequency, the \emph{"finite size effect"} can be safely ignored and the $\tilde{f}_{\mathrm{PT}}$ can be used to estimate PT frequencies.

Finally we discuss the duration and starting time of PT-induced oscillation. Since the PT oscillation is almost a monochromatic oscillation \cite{Zha:2021fbi}, by looking at the optimal frequencies and amplitudes extracted from various neutrino time series, we can estimate the duration and starting time of the PT-induced oscillations. For example, in Hyper-K at 1\,kpc (i.e. Fig.~\ref{fig:faPT1kpcmap}),  the monochromatic region of $\tilde{f}_{\mathrm{PT}}\approx800~\mathrm{Hz}$ starts at $t_{\mathrm{ST}} \approx 1000~\mathrm{ms}$, suggesting that PT-induced oscillation starts around $1000~\mathrm{ms}$ after bounce. The \emph{"cluster"} with almost constant $\tilde{f}_{\mathrm{PT}}$ ends before the time series starting at $1040~\mathrm{ms}$, which indicates that the ending moment of PT-induced oscillations is at around $1040~\mathrm{ms}$. Additionally, the starting time of the oscillation can also be obtained by searching for the burst with highest $\tilde{a}$ values in the map of $\tilde{a}$ distributions. In this way we can measure the oscillation duration and the starting time. Analysis on neutrino signals from Hyper-K with the CCSNe at a distance of 1 kpc approximately reproduces the PT features predicted in the ZOS21-STOS-B145 model. Based on Fig.~\ref{fig:faPT5kpcmap} and Fig.~\ref{fig:faPT10kpcmap}, at larger distances such as 5 and 10\,kpc, we estimate that $t_{\mathrm{ST}}\approx1000~\mathrm{ms}$ and $t_{\mathrm{dur}}\approx30~\mathrm{ms}$. At 1 kpc, the estimated $t_{\mathrm{dur}}$ deviates from the theoretically predicted one by $\sim 10\%$. The duration inferred for source distances of 5 and 10 kpc is shorter than the one inferred at 1 kpc and deviates from the theoretically predicted $t_{\mathrm{dur}}$ by $\sim 30\%$. This is due to the fact that the oscillatory signal tail with lower amplitude is more difficult to distinguished from the shot noise at larger distances. In this work, the resolution of $t_{\mathrm{dur}}$ is 10 ms. A more accurate determination of duration of PT-induced oscillations may be achieved by increasing the resolution of $t_{\mathrm{dur}}$. In IceCube we have $t_{\mathrm{ST}}\approx990~\mathrm{ms}$ and $t_{\mathrm{dur}}\approx50~\mathrm{ms}$ at 5 kpc, $t_{\mathrm{ST}}\approx1000~\mathrm{ms}$ and $t_{\mathrm{dur}}\approx30~\mathrm{ms}$ at 10 kpc. The estimated $t_{\mathrm{dur}}$ on IceCube at 5 kpc agrees well with the theoretical predicted $t_{\mathrm{dur}}$, and has $\sim 30\%$ deviation from theoretical predicted one at 10 kpc.

\begin{table*}
	\caption{\label{tab:hyppara}
	Mean and standard deviation of the parameter distributions for \hk.
	at 10 kpc}
	\begin{ruledtabular}
		\begin{tabular}{ccccc}
		    PT & $f$ Hz &$\delta f / <f>$ Hz&
			$a$  & $\delta a / <a>$  \\
			\hline
			$[940,30]$ ms& 658.5 & 0.22 & 0.099 &0.33
			 \\
			$[940,40]$ ms& 666.7 & 0.22& 0.088 &0.31
			 \\
			$[940,50]$ ms& 653.4 & 0.21& 0.080 &0.24
			 \\
			$[940,60]$ ms& 632.0& 0.18 & 0.086 &0.27
			  \\
			$[940,70]$ ms& 651.6& 0.13 & 0.096 &0.24
			  \\
			$[1010,30]$ ms&695.6 & 0.22 & 0.095 &0.30
			  \\
			$[1010,40]$ ms& 706.5& 0.20 & 0.076 &0.30
			  \\
			$[1010,50]$ ms& 695.4& 0.21 & 0.065 &0.27
			  \\
			$[1010,60]$ ms& 698.8& 0.22 & 0.068 &0.29
			  \\
			$[1010,70]$ ms& 695.1& 0.22 & 0.065 &0.30
			  \\

		\end{tabular}
	\end{ruledtabular}
	
\end{table*}

\begin{table*}
	\caption{\label{tab:hyppara1kpc}
	Mean and standard deviation of the parameter distributions for \hk.
	at 1 kpc}
	\begin{ruledtabular}
		\begin{tabular}{ccccc}
		     PT & $f$ Hz &$\delta f / <f>$ Hz&
			$a$  & $\delta a / <a>$  \\
			\hline
			$[940,30]$ ms& 640.6 & 0.23 & 0.018 &0.63
			 \\
			$[940,40]$ ms& 663.7 & 0.21& 0.014 &0.47
			 \\
			$[940,50]$ ms& 611.1 & 0.21& 0.045 &0.34
			 \\
			$[940,60]$ ms& 560.0& 0.0 & 0.066 &0.07
			  \\
			$[940,70]$ ms& 734.9& 0.01 & 0.073 &0.06
			  \\
			$[1010,30]$ ms&802.1 & 0.008 & 0.048 &0.13
			  \\
			$[1010,40]$ ms& 807.3& 0.012 & 0.033 &0.16
			  \\
			$[1010,50]$ ms& 818.7& 0.006 & 0.032 &0.23
			  \\
			$[1010,60]$ ms& 818.9& 0.006 & 0.030 &0.15
			  \\
			$[1010,70]$ ms& 819.4& 0.007 & 0.029 &0.11
			  \\
			
		\end{tabular}
	\end{ruledtabular}
	
\end{table*}

\subsection{Physics of PT Parameters: Oscillatory Neutrino Signal and Oscillating Proto Compact Stars}\label{sec:viscosity}

In Fig. \ref{fig:neurouosc}, amplitudes of the central density $\tilde{\rho}$ and neutrino luminosity $\tilde{L}_\nu$ oscillation are both shown in the frequency domain. It is interesting to note that as the compactness of the progenitor increases, the PT-induced frequency of both the density and the neutrino luminosity oscillation increases. However, the relationship between progenitor compactness and oscillating PCS frequencies may be heavily model-dependent, and a complete exploration of such a compactness-PCS oscillation frequency relationship is left for future studies. 

Regardless of the underlying progenitor compactness, the frequency of the PT-induced neutrino luminosity oscillations $f_{\rm PT}$ is approximately $f_{\rm PCS}$ and $t_{\rm dur}$ is approximately $\tau_{\rm damp}$, where $f_{\rm PCS}$ is the frequency of the PCS density oscillation and $\tau_{\rm damp}$ is the damping time scale of the PCS. Due to the strong correlation between the oscillatory PT-induced neutrino signals and the oscillating PCS densities explained in section \ref{sec:general}, we can estimate the frequency and damping timescale of PCS oscillations by measuring the $f_{\rm PT}$ and $t_{\rm dur}$ of the PT-induced neutrino signal. With the measured oscillation frequency and duration, we make the suggestion that for the first time the effective bulk viscosity of hadron-quark matter in PCSs could be constrained via the neutrino signal. The damping time of an oscillating spherical star is~\cite{Alford:2019qtm}: 
\begin{equation}\label{eq:taudiss}
    \tau_{\mathrm{damp}}=2\dfrac{\epsilon}{d\epsilon/dt}=\dfrac{2\kappa_{s}^{-1}}{\omega^2\xi},
\end{equation}
where  $\epsilon$ is the energy density of an adiabatic PCS density oscillation, $\xi$ is the PCS effective bulk viscosity, $\omega$ is the oscillation frequency of the PCS density, and $\kappa_s$ is the adiabatic compressibility. Note, the adiabatic and isothermal quantities become equivalent at zero temperature. At finite temperature, because thermal
equilibration is slow in neutrino-transparent nuclear
matter in PCS conditions, we consider adiabatic density oscillations and compressibility here. Readers are referred to \cite{Alford:2019qtm} for more details about isothermal and adiabatic quantities. With $\kappa_s^{-1}=\rho c_S^2$, we have~\cite{Most:2021zvc}:
\begin{equation}\label{eq:taudiss2}
    \tau_{\mathrm{damp}}=\dfrac{2\bar{\rho} c_S^2}{\omega^2\xi}.
\end{equation}

For illustration purpose, let us take the frequency $\omega$ and damping duration $\tau_{damp}$ based on the ZOS21-STOS-B145 model and measured by the PT-meter. As reported in previous section, in this case $\omega=2\pi\times f_{\rm PCS}\approx 2\pi\times800~{\rm Hz}$ and $\tau_{\rm damp}\approx50~{\rm ms}$. Taking a 10 km core (roughly where mixed phase extends to), the averaged density of hadron-quark phase is $\bar{\rho}\approx5.2\times10^{14}$~g~cm$^{-3}\approx1.9\rho_0$ in \cite{Zha:2021fbi}, where $\rho_0$ is the saturation density. The speed of sound of matter with the presence of quark is constrained by $c_S^2\leq 1/3$ at densities high enough that quark matter is nearly perturbative~\cite{Annala:2019puf}. Combined together, we then have:
\begin{equation}
    \xi_{\mathrm{eff}}\lessapprox 2.5\times10^{29}~\mathrm{g}/(\mathrm{cm}~\mathrm{s}).
\end{equation}

Theoretical calculations of the bulk viscosity $\xi_{\rm theo}$ of the hadron-quark mixed phase can be heavily model-dependent because of the unknown surface tension between quark and hadronic matter, and is beyond the scope of this work.  However, it is interesting to compare the effective viscosity $\xi_{\mathrm{eff}}$ extracted from the future measurements of PT-induced neutrino oscillation features with the $\xi_{\mathrm{theo}}$ coming from theoretical calculations of bulk viscosity of hadron-quark matter. Any microphysical viscosity comparable to $\xi_{\mathrm{eff}}$ may affect the dynamics of the star and, hence, the neutrino emission of failing CCSNe. Additionally, an $\xi_{\mathrm{theo}}$ comparable to $\xi_{\mathrm{eff}}$ motivates a self-consistent numerical simulations including physical viscosities.

\section{Summary and discussion}
\label{sec:disc}
In this work we use a novel likelihood method (the PT-meter) to estimate the detectability and the quantitative features of the PT-induced neutrino signals from failing CCSNe. Based on numerical simulations of a failed CCSN including a PT \cite{Zha:2021fbi}, we found the PT-induced neutrino signals from a failing CCSN of a distance within 10 $\mathrm{kpc}$ can be identified in both IceCube and Hyper-K with high statistical confidence. The statistical confidence of the PT detection in various time series of the neutrino signal is quantitatively described using the detection probability ($P_D$) at fixed false alarm rate ($P_{FA}$). We then estimate the frequency, amplitude, and duration of the PT-induced neutrino signals. We further relate these features with the properties of oscillating PCS in failing CCSNe. We show that the effective bulk viscosity of oscillating PCSs can be constrained by the measurement of $f_{\rm PT}$ and $t_{\rm dur}$ in the PT-induced neutrino signal.  

We then discuss the application of PT-meter in more detail. The PT-meter determines if imprints of the PT on the neutrino signals exist by comparing the $\mathcal{L}$ (see eq. \ref{eq:likeR}) from observations with the theoretically predicted PDF of $\mathcal{L}$ from numerical simulations with and without PT activities. This method provides a systematic and quantitative test for theories predicting PT activities. We outline the procedure of applying PT-meter on future observations of neutrino signals from failing CCSNe in the following:
\begin{itemize}
\item Preparation I --- Fourier transform neutrino signals with various $\tau$s and $t_{ST}$s from two models to be compared with observations, using eq. \ref{eq:power1}. The chosen $\tau$s and $t_{ST}$s are illustrated in Fig. \ref{fig:PDmap} and Fig. \ref{fig:PDmapIce}. 
\item Preparation II --- Simulate the realistic Fourier transformed neutrino signals by taking many realizations of these two models and therefore sampling the distribution from the shot noise. Calculate the $\mathcal{L}$s of these realizations using eq. \ref{eq:likeR}, and determine the PDFs of $\mathcal{L}$. 
\item Preparation III --- Given the PDFs of $\mathcal{L}$ based on two models, a receiver operating characteristic (ROC) curve ($P_D$ as a function of $P_{FA}$) can be determined. Find the optimal $\tau$ and $t_{ST}$ of the neutrino signal, that give the best ROC curve. Determine the $\mathcal{L}_{thr}$ that gives low $P_{FA}$ and high $P_D$ based on the PDFs of $\mathcal{L}$ in the neutrino signals with optimal $\tau$ and $t_{ST}$.
\item Determination of existence of PT-induced oscillations --- Fourier transform the observed neutrino signal with $\tau$s and $t_{ST}$s defined in step one, calculate the $\mathcal{L}_{obs}$s based on observations using eq. \ref{eq:likeR}. Compare $\mathcal{L}_{obs}$ with $\mathcal{L}_{thr}$ in neutrino signals with optimal $\tau$ and $t_{ST}$ found in step three. If $\mathcal{L}_{obs}>\mathcal{L}_{thr}$, model with PT is favored given the observation, with $P_{D}$ and $P_{FA}$ corresponding to the chosen $\mathcal{L}_{thr}$. 
\item Determination of features of PT-induced oscillations --- Based on observed neutrino signals, produce a map of $f_{PT}$ in various $\tau$s and $t_{ST}$s similar as Fig. \ref{fig:faPT1kpcmap}. The starting time and the duration of the PT can be found by localizing the monochromatic cluster of $f_{PT}$ on this map, and the characteristic frequency of PT is the averaged frequency in the cluster of this map, as discussed in section \ref{sec:parameter}. The relative uncertainties of observed PT features is approximately to be the same as the ones obtained from simulations at the same distance (see section \ref{sec:parameter}). 
\end{itemize}

Note, in this paper introducing the prototype of the PT-meter, a core-collapse simulation of a 17 solar-mass star by Zha \emph{et al}. (2021) \cite{Zha:2021fbi}
is used as representative of a future PT-induced oscillatory signal. In reality, the strength and the starting time of the PT-induced oscillatory signals may not be the same as the prediction of this model. If the real PT-induced signal starts much earlier than the model predicts, or the real PT-induced signal has much smaller oscillation amplitude, the PT-meter may "miss" the signal and give a false negative result. The occurrence of false negative result may come with two inconsistent behaviors of the PT-meter: (1) the $\mathcal{L}_{obs}$ is obviously higher than the $\mathcal{L}$ of the chosen PT model in a region before or after the predicted PT-induced imprints; (2) the map of $f_{PT}$ similar as that in Fig. \ref{fig:faPT1kpcmap} but produced based on real signal has an obvious monochromatic cluster, while the PT-meter gives a negative result of PT existence. The inconsistent behaviors of PT-meter suggest that models with other $t_{ST}$ and PT-induced oscillation amplitude may need to be considered and be compared with observations. In section \ref{sec:result}, we discuss the most basic situation, where the PT-meter is applied to distinguish two models with and without PT activities by \emph{Zha et al.}. Note that, the impact of equation
of states, neutrino transport schemes, and dimensionality of the simulations on PT-induced
neutrino oscillatory signals have not been systematically studied. And some of the aforementioned factors may induce quantitative and even qualitative differences on the PT-induced neutrino signals. We discuss some dependencies on the equation of state and dimensionality in more detail in the appendix. 
Consequently, it will be interesting to use this method to compare multiple numerical simulations of failing CCSNe. In this way one can construct a global comparison of numerical
simulations with PT activities. 

Although the phase-transition-induced PCS bounce does not always induce the quasi-periodic kilo Hertz neutrino signals, it might be true that these late-time oscillating neutrino signals can only be produced immediately after the PT-induced core bounce. Consequently, the detection of kilo Hertz oscillating neutrino signals complements the detection of phase transitions using the second neutrino burst discussed in Ref. \cite{Sagert:2008ka,Fischer:2017lag}. More importantly, the combination of the detection of the second neutrino burst with the detection or non-detection of quasi-periodic neutrino signals may provide additional information of the underlying hybrid EoS as well as the rotation speed of the PCS. If the CCSN is non-rotating and one observes the quasi-periodic oscillations of neutrino signals, the frequency as well as the damping time provides a unique way of estimating the effective viscosity of PCS, as discussed in this work.      
 
For the first time by using the PT-meter, we have a systematic and quantitative statistical method to compare the observations to numerical simulations with and without PT-induced neutrino signals. Additionally, if a time series is identified to have PT-induced oscillation with high statistical confidence level (namely, high $P_D$), the optimal parameters of the template characterizing the PT-induced oscillations can be used to constrain the effective bulk viscosity in protocompact stars, as discussed in section \ref{sec:viscosity}. Finally, similar to the SASI~\cite{Tamborra:2013laa}, the PT leaves imprints on both the neutrino and the GW signals of failing CCSNe. In the future, it could be interesting to combine the neutrino analysis with a GW analysis to study PT activities.

\subsubsection*{Acknowledgments}
ZL thanks A. Haber, C. Lunardini, E.R. Most, D. Vartanyan and A. Burrows for helpful discussions. ZL acknowledges funding from the NSF Grant PHY
21-16686, PHY-1554876 and funding from DOE Scidac Grant DE-SC0018232. AWS was supported by NSF PHY
21-16686 and PHY-190490. This work is supported by the Swedish Research Council (Project No. 2020-00452). S.Z. is supported by the National Natural
Science Foundation of China (Grant Nos. 12288102, 12090040, 12090043) and the International
Centre of Supernovae, Yunnan Key Laboratory (No. 202302AN360001). Some simulations were enabled by resources provided by the Swedish National Infrastructure for Computing (SNIC) at NSC partially funded by the Swedish Research Council through grant agreement No. 2018-05973. 
\clearpage
\newpage
\appendix
\section{EoS-, Rotation-Dependent PT-induced neutrino signals}\label{appendix1}
In the last several decades, the measurements from heavy ion collision experiments have strengthened the constraints on EoSs \cite{Danielewicz:2002pu, Bastian:2020unt, Huth:2021bsp}, and several new types of hybrid EoSs \cite{Alvarez-Castillo:2016oln,Kaltenborn:2017hus, Bastian:2020unt} have widely been used in recent simulations of CCSNe considering the existence of quark matter. It is also worth mentioning that the rotation of compact stars may heavily influence the PT-induced core collapse. Here, we discuss some dependencies of the PT-induced neutrino signals on the EoS and on the PCS rotation rate. 

\begin{figure}
    \centering
    \includegraphics[width=0.45\textwidth]{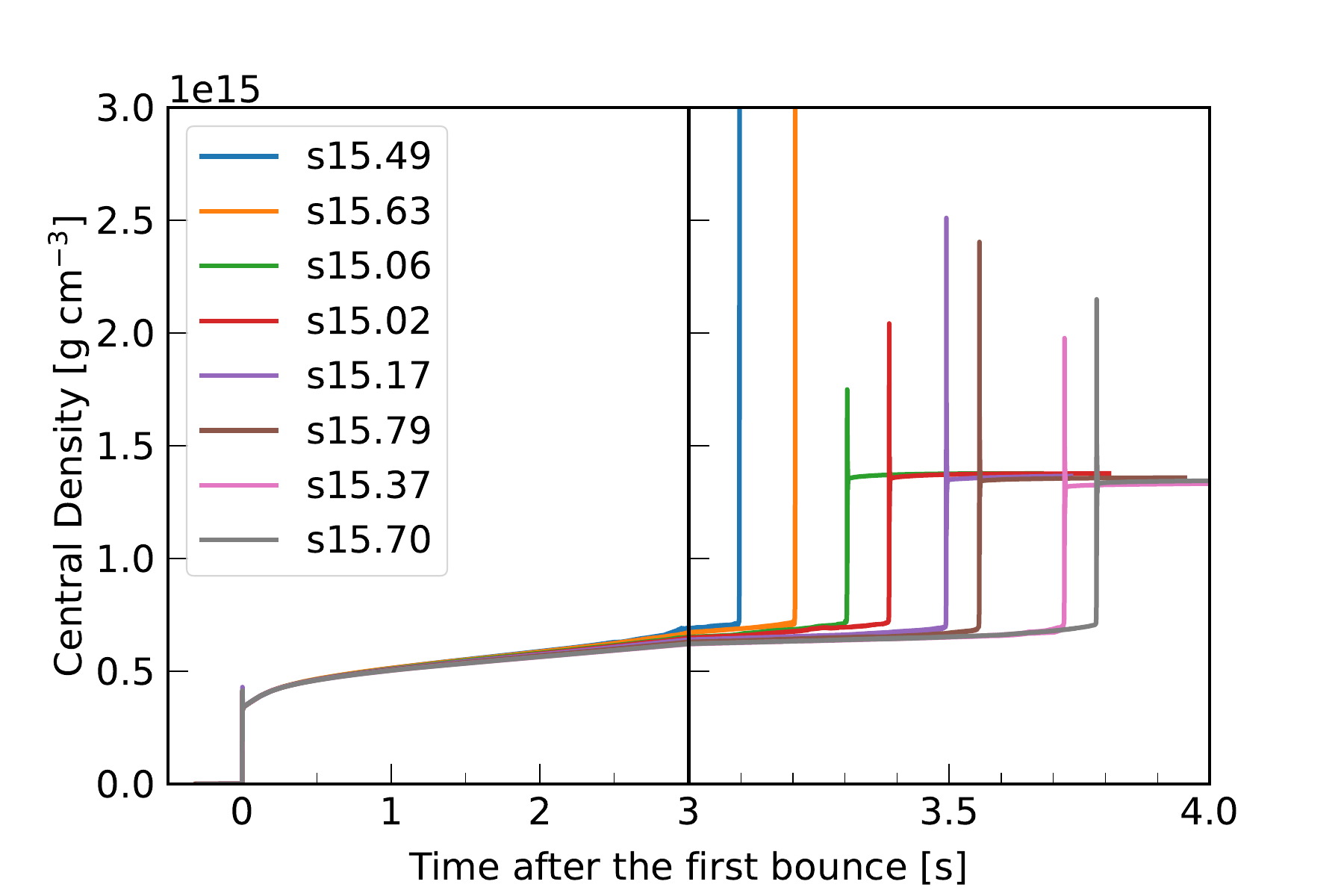}
    \includegraphics[width=0.45\textwidth]{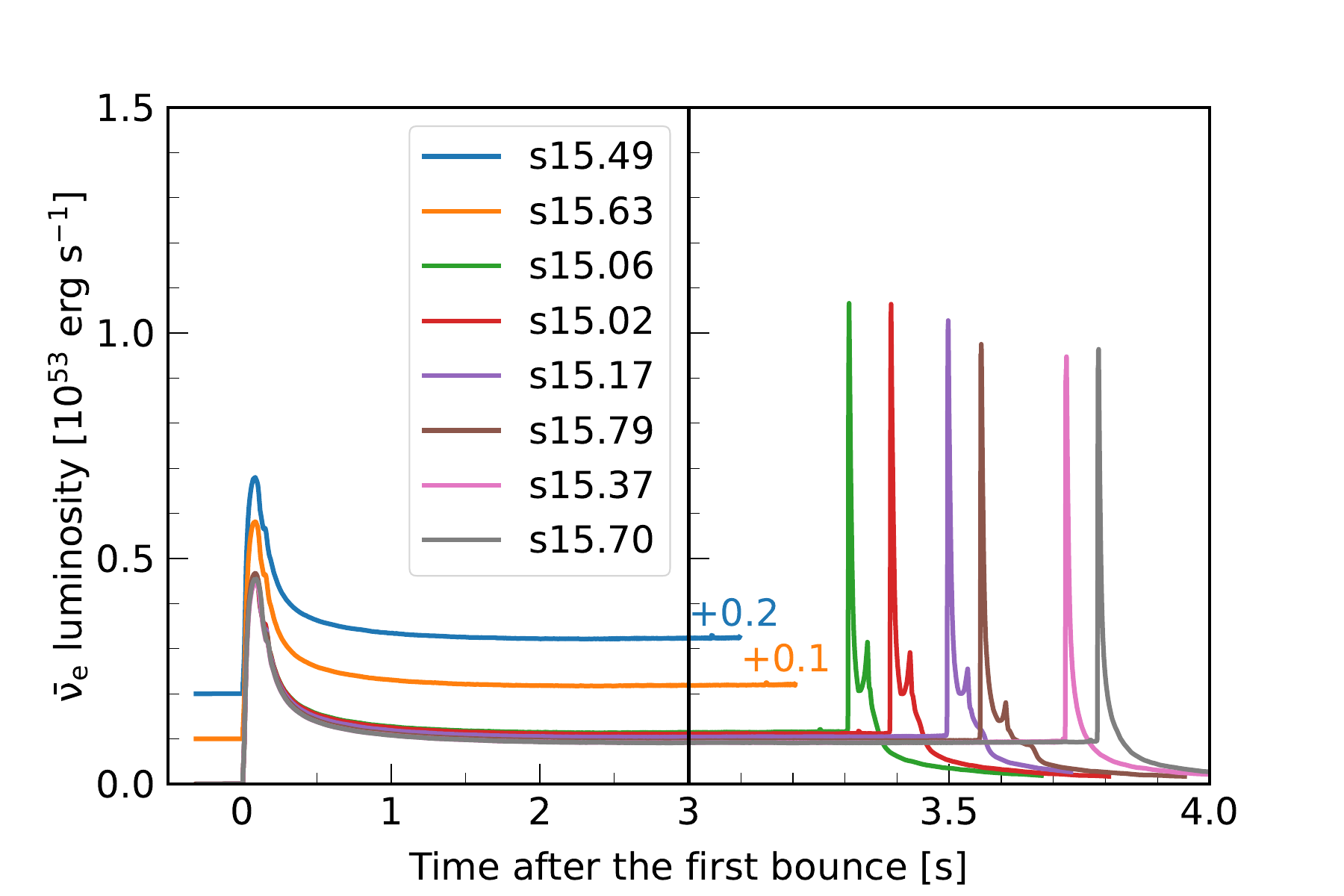}
    \caption{Top panel: time evolution of the central density in eight selected non-rotating 1D models with progenitor masses varying from 15.02\,$M_\odot$ to 15.79\,$M_\odot$. Bottom panel: Time evolution of the neutrino luminosity in 8 selected non-rotating 1D models with progenitor mass varying from 15.02\,$M_\odot$ to 15.79\,$M_\odot$. Note that in models with progenitor mass of 15.49\,$M_\odot$ and 15.63\,$M_\odot$, the PCS immediately collapse to black holes upon reaching the PT and the neutrino luminosities terminate at $\sim 3.1 \mathrm{s}$ and $\sim 3.2 \mathrm{s}$ after the first bounce.}
    \label{fig:central density and luminosity}
\end{figure}

We first studied eight additional one-dimensional, non-rotating CCSNe simulations with progenitor masses between $15\,M_\odot$ and $16\,M_\odot$ using the RDF1.4 EoS \cite{Bastian:2020unt}. Among these eight simulations, two immediately collapse to black holes upon reaching the PT and six of them explode following the PT. In the non-rotating CCSNe models using RDF1.4 EoS, the central densities of the PCSs after the second bounce are much higher than the ones based on our simulation using STOS-B145 in Fig. \ref{fig:sim}.  Additionally, the shock wave does propagate outward without being stalled near the neutrinosphere for a long time after the second collapse as is the case in the main model studied here. We further observe that there are no obvious oscillations in either the central densities or the neutrino luminosities based on these eight models. The central densities as well as the neutrino luminosities of these eight additional models are presented in Fig.\ref{fig:central density and luminosity}. This, however, does not exclude the possibilities of observing quasi-periodic oscillations in non-rotating models using state-of-the-art hybrid EoS different from RDF1.4 \cite{Jakobus:2022ucs}. The relationship between the EoS and the neutron star quasi-periodic oscillations was systematically studied in Ref.~\cite{Sun:2021cez,Kunjipurayil:2022zah,Zhao:2022tcw}. 
However, the relationship between the EoS and the oscillations in PCSs studied here is less clear. The properties of the EoS may influence the expansion of the shock wave, and further influence the occurrence probability of oscillations in PCS. It would be interesting to systematically study the effect of the EoS on PCS oscillations in CCSNe in future investigations.

In Ref. \cite{Zha:2022pnc}, rotating models using the RDF1.9 EoS were studied. The central density of PCS at the second bounce decreases as the rotation speed increases because of the added rotational support. The quasi-periodic radial oscillation mode (with $l$=0) is still severely damped in the center of the PCS but becomes less damped at larger radius \cite{Zha:2022pnc}. Correspondingly, oscillations in both the PCS radii and in the neutrino luminosities are observed \cite{Zha:2022pnc}. Although the excitation mechanism of PCS oscillation modes in rotating models could be different from that in non-rotating ones, the frequency as well as the amplitude of the oscillations in the rotating models observed in Ref. \cite{Zha:2022pnc}, is the same order of magnitude as the ones from the non-rotating models using STOS-B145 explored here. It suggests that the PT-meter method proposed in this work, is also suitable to study the quasi-periodic neutrino signals from rotating CCSNe.  

\clearpage
\newpage
\bibliography{b}


\end{document}